\documentclass[useAMS,fleqn,usenatbib]{mnras}
\usepackage[utf8]{inputenc}
\usepackage{amssymb,amsmath}
\usepackage{graphicx}
\usepackage[dvipsnames]{xcolor}
\hypersetup{colorlinks=true,allcolors=teal}
\usepackage[flushleft]{threeparttable}

\newcommand{\be}{\begin{equation}}
\newcommand{\ee}{\end{equation}}
\def\bear#1\ear{\begin{align}#1\end{align}}
\newcommand{\nline}{\notag \\}
\newcommand{\f}{\frac}

\newcommand{\Msun}{\mathrm{M}_{\odot}}

\newcommand{\eqn}[1]{equation (\ref{#1})}

\newcommand{\secn}[1]{Section \ref{#1}}
\newcommand{\appndx}[1]{Appendix~\ref{#1}}
\newcommand{\fig}[1]{Fig. \ref{#1}}
\newcommand{\tab}[1]{Table \ref{#1}}

\usepackage{newtxtext,newtxmath}

\usepackage[T1]{fontenc}

\DeclareRobustCommand{\VAN}[3]{#2}
\let\VANthebibliography\thebibliography
\def\thebibliography{\DeclareRobustCommand{\VAN}[3]{##3}\VANthebibliography}

\usepackage{graphicx}	
\usepackage{amsmath}	
\usepackage{subfig}

\title[Reionization and thermal history using SCRIPT]{Constraining the reionization and thermal history of the Universe using a semi-numerical photon-conserving code SCRIPT}

\author[Maity \& Choudhury]{
Barun Maity$^{1}$\thanks{E-mail: bmaity@ncra.tifr.res.in}
and Tirthankar Roy Choudhury$^{1}$\\
$^{1}$National Centre for Radio Astrophysics, TIFR, Pune University Campus, Post Bag 3, Pune 411 007, India}
\date{Accepted XXX. Received YYY; in original form ZZZ}

\pubyear{2021}

\begin{document}
\label{firstpage}
\pagerange{\pageref{firstpage}--\pageref{lastpage}}
\maketitle

\begin{abstract}
Given that the reionization history of cosmic hydrogen is yet to be stringently constrained, it is worth checking the prospects of doing so using physically motivated models and available observational data.
For this purpose, we use an extended version of the explicitly photon-conserving semi-numerical model of reionization, \texttt{SCRIPT}, which also includes thermal evolution of the intergalactic medium (IGM).
The model incorporates the effects of inhomogeneous recombination and radiative feedback self-consistently and is characterized by five free parameters (two for the redshift-dependent ionization efficiency, two for the ionizing escape fraction, and another for reionization temperature increment).  
We constrain these free parameters by simultaneously matching with various observational probes, e.g., estimates of the ionized hydrogen fraction, the CMB scattering optical depth and the galaxy UV luminosity function. 
In addition, we include the low-density IGM temperature measurements obtained from Lyman-$\alpha$ absorption spectra at $z \sim 5.5$, a probe not commonly used for Bayesian analysis of reionization parameters.
We find that the interplay of the various data sets, particularly inclusion of the temperature data, leads to tightening of the parameter constraints. 
Our default models prefer a late end of reionization (at $z \lesssim 6$), in agreement with other recent studies. 
We can also derive constraints on the duration of reionization, $\Delta z=1.81^{+0.51}_{-0.67}$ and the midpoint of reionization, $z_{\mathrm{mid}}=7.0^{+0.30}_{-0.40}$. 
The constraints can be further tightened by including other available and upcoming data sets.

\end{abstract}




\section{Introduction}
\label{sec:introduction}

The epoch of reionization, the cosmic age when the radiation from the first stars ionized hydrogen atoms in the surrounding intergalactic medium (IGM), is perhaps one of the less understood phases in the history of the universe. Developing physical understanding of reionization involves rather complex non-linear astrophysics at small scales on one hand and cosmological evolution at very large scales on the other. This turns out to be a challenging task. What makes the study of this epoch so interesting is the availability of observational data over a wide range of the wavebands. In the most recent times, we have gained information on the ionization history of the universe from the Cosmic Microwave Background (CMB) \citep{2020A&A...641A...6P}, fluctuations in the opacity of the Lyman-$\alpha$ (Ly$\alpha$) absorption spectra of quasars at $z \lesssim 6$ \citep{2015MNRAS.447.3402B,2018MNRAS.479.1055B,2018ApJ...864...53E,2019ApJ...881...23E,2020ApJ...904...26Y,2021arXiv210803699B} and the properties of the Ly$\alpha$ emitters \citep{2014ApJ...797...16K,2017ApJ...844...85O,2017ApJ...842L..22Z,2018PASJ...70S..16K,2018ApJ...867...46I,2018PASJ...70S..13O}. To complement these, we can probe the sources of reionization using the ultra-violet luminosity function (UVLF) of high-redshift galaxies \citep{2015ApJ...810...71F,2015ApJ...800...18A,2018MNRAS.479.5184A,2015ApJ...803...34B,2017ApJ...843..129B,2021AJ....162...47B}.
 The ionization history is also closely coupled to the thermal evolution of the IGM which can be studied through the temperature measurements at $z \gtrsim 5$ obtained from the Ly$\alpha$ absorption spectra \citep{2019ApJ...872...13W,2020MNRAS.494.5091G}.

A possible approach to model the epoch of reionization and compare with observations is to use the semi-numerical simulations. While these models are not as elaborate as the full hydrodynamic/radiative transfer simulations, they have the advantage of being computationally efficient and thus useful for parameter space exploration. In case one needs to compare the theoretical predictions with a wide variety of observations simultaneously and constrain the full possible range of allowed histories, the semi-numerical models provide an ideal option.

A major challenge for carrying out parameter space exploration using these numerically efficient models is to maintain the efficiency of the models while incorporating various astrophysical processes during reionization which are intrinsically inhomogeneous in nature.  While early semi-analytical models \citep{2003ApJ...586..693W,2010MNRAS.408...57P,2006MNRAS.371L..55C,2011MNRAS.413.1569M,2012MNRAS.419.1480M,2015MNRAS.454L..76M} have been reasonably successful in constraining the globally averaged quantities, these do not model the inhomogeneities in the ionization or thermal fields during reionization, both of which can play crucial roles in interpreting the data. There are several semi-numerical models which are able to account for these fluctuations in the ionization field based on the excursion set approach, e.g., \texttt{21cmFAST} \citep{2007ApJ...669..663M,2011MNRAS.411..955M} and \texttt{SIMFAST21} \citep{2010MNRAS.406.2421S}. A slightly different and more simplistic approach has been implemented in \texttt{zreion} \citep{2013ApJ...776...81B} which assumes the reionization redshift field to be a biased tracer of the density field. For further insights on the coupling between reionization and galaxy formation, one needs to incorporate other important physical effects like the inhomogeneous recombinations \citep{2009MNRAS.394..960C,2013MNRAS.432.3340S}, and radiative feedback which suppresses  the star formation inside low mass haloes \citep{2000ApJ...542..535G,2007MNRAS.376..534I,2013MNRAS.432.3340S,2021MNRAS.503.3698H}. Although it is possible to include relatively detailed physical modelling within the semi-numerical reionization framework, e.g., \texttt{DRAGONS} \citep{2016MNRAS.462..250M,2016MNRAS.462..804G}, \texttt{ARTIST} \citep{2019MNRAS.489.5594M} and \texttt{ASTRAEUS} \citep{2021MNRAS.503.3698H}, but those usually come at the cost of efficiency. Hence most of the semi-numerical models assume simplified parametrization for galaxy properties,  often sufficient to solve the problem. Ultimately, the models are used for simultaneously computing various observables like the UVLFs, Ly$\alpha$ absorption spectra and the 21~cm power spectra \citep{2015MNRAS.449.4246G,2017MNRAS.468..122H} which can then be compared with the relevant observational data. These models can be further utilized to couple with advanced Bayesian statistical tools for inference studies \citep{2015MNRAS.449.4246G,2017MNRAS.472.2651G,2018MNRAS.477.3217G,2019MNRAS.484..933P,2021MNRAS.506.2390Q}. 

Most of the semi-numerical models of reionization that explore the unknown parameter space using Bayesian techniques do not include the observables related to the thermal evolution of the IGM as a probe of reionization. The thermal evolution as a probe of reionization has been proposed in several earlier studies \citep{1994MNRAS.266..343M,2002ApJ...567L.103T,2003ApJ...596....9H,2010MNRAS.406..612B} and there also exist semi-analytical models for probing thermal evolution during  reionization, mainly that of singly-ionized helium \citep{1997MNRAS.292...27H,2016MNRAS.456...47M,2016MNRAS.460.1885U}. Ideally, the thermal evolution of the medium should to be coupled to the reionization modelling with the hope that it may lead to more stringent constraints of the reionization parameters.

In our previous work \citep[][hereafter Paper I]{2022MNRAS.511.2239M}, we introduced a semi-numerical model for studying the ionization and thermal history of the universe. The model was based on an explicitly photon-conserving code for generating the ionization field given the sources of reionization, named \texttt{SCRIPT} \citep{2018MNRAS.481.3821C}. The code was coupled to the equation for temperature evolution, which allowed us to account for the effect of inhomogeneous reionization on the temperature field. We included several physical effects which could be important during reionization, namely, the inhomogeneous recombinations and radiative feedback. In Paper I, we demonstrated the usefulness of our code by computing several observables for a default reionization model, and also studied a few variants.

A natural follow up of our previous work would be to carry out a full exploration of the parameter space. This would allow us to understand degeneracies between different free parameters and how they affect the allowed range of reionization histories. In this work, we use the model presented in Paper I to perform a detailed study of the parameter estimation using a Monte Carlo Markov Chain (MCMC) based Bayesian analysis and provide the constraints on ionization and thermal evolution of the IGM. This would allow us to not only understand the importance of the presently available data in constraining reionization but also to appreciate the abilities of the upcoming experiments. 

 The paper is organized as follows: In \secn{sec:brief_model}, we give a brief introduction to our models and different parameters involved. We discuss about the  different observational constraints in \secn{sec:obs_cons}. In \secn{sec:parms_stat}, we give the basic formalisms for the parameter statistics. Then we discuss the results in \secn{sec:results}. Lastly, we summarize and conclude this work in \secn{sec:conc}. In this paper, the assumed cosmological parameters are $\Omega_M$ = 0.308, $\Omega_{\Lambda}$ = 0.691 $\Omega_b$ = 0.0482, $h$ = 0.678, $\sigma_8$ = 0.829 and $n_s$ = 0.961 \citep{2016A&A...594A..13P}.

\section{Theoretical model}
\label{sec:brief_model}

The theoretical model to generate the ionization and thermal history of the universe used in this work is identical to that in Paper I \citep{2022MNRAS.511.2239M}. We summarize the main features of the model and refer the reader to Paper I for more details.

\begin{itemize}

\item The core component of the model is the explicitly photon-conserving semi-numerical code \texttt{SCRIPT} which allows one to generate the ionization fields \citep{2018MNRAS.481.3821C} in a cosmologically representative simulation volume divided into grid cells. The main advantage of this code over others is that it leads to a large-scale field that is numerically convergent with respect to the resolution of the simulation. The generation of the ionization field requires two other fields, namely, those corresponding to the density and the collapsed haloes capable of hosting ionizing sources. Since we are interested in the properties of the IGM only at large scales, it is more efficient to use a 2LPT code to generate the density field \citep{2011MNRAS.415.2101H}\footnote{\url{https://www-n.oca.eu/ohahn/MUSIC/}}. The halo field is obtained using a sub-grid prescription based on the conditional ellipsoidal mass function \citep{2002MNRAS.329...61S}. In this work, we use a grid cell resolution of $16h^{-1}\mathrm{cMpc}$ with the simulation box size of $256h^{-1}\mathrm{cMpc}$ which is adequate for the observables we calculate in the work.

\item The ionization field is generated using the explicitly photon-conserving algorithm of \citet{2018MNRAS.481.3821C}. The amount of photons produced by the haloes is parametrized by the reionization efficiency parameter $\zeta(M_h, z)$ which essentially is the number of ionizing photons escaping into the IGM per hydrogen atom in collapsed haloes. This efficiency parameter, in general, can depend on redshift $z$ and the halo mass $M_h$. As introduced in Paper I, our algorithm for generating the ionization field accounts for the number of recombinations in each grid cell of the simulation integrated over its reionization history. The recombinations are controlled by the small-scale inhomogeneities in the density field which are not resolved in our simulations. We parametrize these sub-grid fluctuations in terms of a globally averaged clumping factor $C_{\mathrm{HII}}$.

\item Since the recombination rate depends on the temperature of the medium, we complement the ionization history by solving also for the thermal history of each grid cell in the box. We ensure that the thermal evolution of a cell is coupled to its ionization history, i.e., our code automatically accounts for the effect of spatially inhomogeneous reionization on the temperature. As a region gets ionized for the first time, we assume that its temperature increments by a value $T_{\mathrm{re}}$, which is a free parameter. This temperature is referred to as the reionization temperature \citep{1997MNRAS.292...27H,2009ApJ...701...94F,2018MNRAS.477.5501K,2022MNRAS.511.2239M}.

\item Our method also includes radiative feedback suppressing the production of ionizing photons in haloes where the gas is heated up. In Paper I, we introduced several methods for implementing the feedback. In this work, our default scheme for implementing feedback is the step feedback model. In this case, the gas fraction retained inside a halo affected by feedback is assumed to be zero for a halo of mass smaller than
\be
M_{\mathrm{min}} = \mathrm{Max} \left[M_{\mathrm{cool}}, M_{J}\right].
\ee
and unity otherwise. In the above equation, $M_{\mathrm{cool}}$ is the minimum mass of haloes where the gas can cool via atomic transitions and form stars, given by \citep{2001PhR...349..125B,2013MNRAS.432.3340S}
\begin{equation}
\label{eq:Mcool}
M_{\mathrm{cool}} = 10^8 \Msun \left(\frac{10}{1+z}\right)^{3/2}.
\end{equation}
The Jeans mass $M_J$ at virial overdensity determines the minimum halo mass that can form stars in the feedback affected regions and is given by \citep{2021MNRAS.503.3698H}
\begin{equation}\label{eq:M_J}
M_{J} = \frac{3.13 \times 10^{10} h^{-1} \Msun}{\Omega_m^{1/2}~(1+z)^{3/2}~\sqrt{18\pi^2}} ~\mu^{-3/2}~ \left(\frac{T_{\mathrm{HII}}}{10^4\mathrm{K}}\right)^{3/2},
\end{equation}
where $\mu$ is the mean molecular weight (assumed to be 0.6, appropriate for ionized hydrogen and singly ionized helium) and $T_{\mathrm{HII}}$ is the temperature of the ionized regions in the grid cell under consideration.

In addition to the step feedback, we also consider a variant of the method where the $M_h$-dependence of the feedback is gradual. In this case, the gas fraction retained inside a feedback affected halo does not suffer a sharp step-like cut-off but has a form \citep{2013MNRAS.432.3340S,2019MNRAS.482L..19C,2022MNRAS.511.2239M}
\begin{equation}\label{eq:fg_gradual}
 f_{g}(M_h) = 2^{-M_{J} / M_h} = \exp\left(-\f{M_{J}}{1.44 M_h}\right),
\end{equation}

\end{itemize}

The model can be used to compute several physical quantities and observables of interest. The evolution of the ionization field provides the global reionization history in the form of either the mass-averaged ionized fraction $Q^M_{\mathrm{HII}}(z)$ or the volume-averaged one $Q^V_{\mathrm{HII}}(z)$. The Thomson scattering optical depth of the CMB photons can be computed for a particular reionization history as
\be
\label{eq:tau_eq}
\tau_{e} =\chi_{\mathrm{He}}\sigma_{\mathrm{T}}c\int_{0}^{z_{\mathrm{LSS}}}\frac{Q_{\mathrm{HII}}^{M}(z)n_{\mathrm{H}}(z)(1+z)^2dz}{H(z)},
\ee
where $\sigma_T$ is the Thomson scattering cross section, $\chi_{\mathrm{He}}$ is the contribution of singly ionized helium to the free electron density, $c$ is the velocity of light, $n_H(z)$ is the comoving number density of hydrogen and $H(z)$ is the Hubble parameter at redshift $z$. The integration formally extends to the redshift of last scattering $z_{\mathrm{LSS}}$ although in practice the contribution is essentially limited to the start of reionization.

In this work, we define several quantities which are useful in quantifying the reionization evolution. One of those is the reionization duration $\Delta z$ defined as the redshift interval between which $Q^M_{\mathrm{HII}}$ evolves from 25\% to 75\% \citep{2021MNRAS.500..232P,2021MNRAS.501L...7C}. The other quantities are the midpoint of reionization $z_{\mathrm{mid}}$ defined as the redshift where $Q^M_{\mathrm{HII}} = 0.5$  and $z_{\mathrm{75}}$ defined as the redshift where $Q^M_{\mathrm{HII}} = 0.75$.

Our model can be extended to compute the temperature-density relation of the low-density IGM, an observable relevant to the quasar absorption spectra. For this purpose, we track the thermal evolution of the representative subgrid elements inside our rather coarse-resolution grid cells. These subgrid density elements are distributed assuming a lognormal distribution with a mean same as the cell overdensity and a variance corresponding to the Jeans scale at the initial redshift. The exact form of the distribution is not important for the calculations as long as the elements are centred around the mean density of the parent cell. These elements act as reprsentatives of the low-density IGM. We then solve for the temperature of these subgrid elements taking into account all the relevant physical processes, including patchy reionization \citep[for details, see section 4.1 of][]{2022MNRAS.511.2239M}. This allows us to probe the correlation between the temperature and density at redshifts of interest. In particular, we assume a power-law relation $T(\Delta) = T_0 \Delta^{\gamma-1}$ and obtain the parameters $T_0$ and $\gamma$ by fitting the model outputs. The values of $T_0$ and $\gamma$ can be compared with observations.

We also compute the galaxy UV luminosity function (UVLF) by relating the ionizing properties of the sources of reionization (i.e., galaxies) to their UV properties. The ionization field is sensitive only to the ionization efficiency $\zeta(M_h, z)$ which is a multiplicative combination of the star-forming efficiency $f_*(M_h, z)$ and the escape fraction $f_{\mathrm{esc}}(M_h, z)$ of ionizing photons. However, the UVLF is sensitive only to $f_*$, hence the computation requires the additional knowledge of $f_{\mathrm{esc}}$ (or equivalently $f_*$). Another ingredient needed to compute the UVLF is the relation between ionizing and UV properties of the stellar population. As in Paper I, we assume parameters consistent with galaxies having a Salpeter IMF and metallicity $Z \approx 0.1 Z_{\odot} = 0.002$ \citep[estimated using, e.g., \texttt{STARBURST99},][]{1999ApJS..123....3L}. Note that these parameters characterizing the stellar populations are degenerate with the escape fraction.

From the above discussion, it is clear that our model contains several free parameters, we summarize them below:

\begin{itemize}

\item The first parameter of interest is ionization efficiency $\zeta$ which we assume to be independent of the halo mass. We take a redshift dependent form of ionization efficiency \citep{2015ApJ...813...54T,2016MNRAS.460..417S,2020MNRAS.495.3065D,2021MNRAS.501L...7C,2022MNRAS.511.2239M} as a simple power-law
\be\label{eq:zeta_eq}
    \zeta(z) = \zeta_0\left(\frac{10}{1+z}\right)^\alpha,
\ee
where $\zeta_0$ is the ionization efficiency at redshift 9 and $\alpha$ is the slope of the power-law. We use $\log \zeta_0$ and $\alpha$ as free parameters in our models. 

\item The reionization temperature $T_{\mathrm{re}}$, essential in the modelling of the IGM temperature evolution, recombination and radiative feedback, is also a free parameter. In general, this temperature depends on the velocities of the ionization fronts and the spectra of the ionizing sources, as found in the high-resolution simulations of \citet{2019ApJ...874..154D}, and hence may evolve as the reionization progresses. The computation of ionization front velocities require much higher resolutions than what we use, and hence is beyond the scope of our models. Hence we take it to be a constant independent of the redshift.

\item Computation of the UVLFs requires the knowledge of $f_{\mathrm{esc}}(M_h, z)$. Since we will be comparing the model with UVLF data only at redshifts $z = 6$ and $7$, the value of $f_{\mathrm{esc}}$ needs to be computed only at these two epochs. We assume that the evolution in $f_{\mathrm{esc}}$ from $z = 7$ to $6$ can be neglected and take it to be a solely $M_h$-dependent quantity \citep{2015MNRAS.451.2544P,2016ApJ...833...84X,2020MNRAS.498.2001M}. We hence define it as
\be\label{eq:esc_eq}
     f_{\mathrm{esc}} = f_{\mathrm{esc}}^0\left(\frac{M_h}{10^9M_{\odot}}\right)^\beta,
\ee
where $f_{\mathrm{esc}}^0$ and $\beta$ are free parameters. Since the ionization efficiency $\zeta \propto f_*~f_{\mathrm{esc}}$ is taken to be mass-independent, the mass dependence of star formation efficiency ($f_*$) turns out to be such that it exactly compensates for the mass dependence of ($f_{\mathrm{esc}}$). In other words, our assumptions automatically lead to a mass dependence of $f_* \propto M_h^{-\beta}$. There is no strong physical reason for this choice, however it is consistent with the parameter estimates obtained by earlier studies \citep{2019MNRAS.484..933P,2021MNRAS.506.2390Q}.

\item For our default runs, we fix the value of the clumping factor $C_{\mathrm{HII}}$ to 3, a choice that is in agreement with radiative transfer simulations \citep{2020ApJ...898..149D}. However we also explore the case where $C_{\mathrm{HII}}$ is treated as a free parameter.

\end{itemize}

So, we have five free parameters $\{\log \zeta_0, \alpha, T_{\mathrm{re}}, f_{\mathrm{esc}}^0, \beta\}$ for our default run. As one variant of the default, we check the constraints where the clumping factor ($C_{\mathrm{HII}}$) is kept as a free parameter.

\section{Observational data sets}

\label{sec:obs_cons}

We use the different available data sets to constrain the model parameters and the corresponding ionization and thermal histories. In this section, we list the observational constraints which are used in this study.

\begin{enumerate}

\item An important constraint on reionization history comes from the CMB scattering optical depth ($\tau_e$). We use the latest Planck measurement \citep{2020A&A...641A...6P} of $\tau_e = 0.054 \pm 0.007$ for our likelihood analysis.

\item 
We use model-independent constraints on the ionization fraction obtained from dark pixel fraction in quasar spectra \citep{2015MNRAS.447..499M}. These give us the upper limit on ionization fractions at relatively lower redshifts ($z \lesssim 6$).

\item We require reionization to be complete (i.e., $Q^M_{\mathrm{HII}}$ should become unity) at $z \geq 5.3$. This limit is motivated by observations of the Ly$\alpha$ optical depth from distant quasars \citep{2018MNRAS.479.1055B,2017ApJ...840...24E,2018ApJ...864...53E,2021ApJ...923..223Z,2021arXiv210913170C} and theoretical models \citep{2019MNRAS.485L..24K,2021MNRAS.501.5782C,2020MNRAS.494.3080N,2020MNRAS.497..906K,2021arXiv210803699B}.

\item From optical studies we have the data on galaxy UVLFs at different redshifts \citep{2015ApJ...803...34B,2017ApJ...843..129B}. The faint end of the UVLFs are sensitive to the feedback and these can be very useful in constraining the heating associated with reionization. We use the constraints from UVLFs at redshifts 6 and 7 when the feedback is supposed to be effective \citep{2015ApJ...803...34B,2017ApJ...843..129B}. 

\item Lastly, we use temperature at low density IGM as an additional probe of reionization history. There exist recent estimates of the $T-\Delta$ power-law relation, parametrized by $T_0$ and $\gamma$,  at $z = 5.4, 5.6$ and $5.8$ \citep{2020MNRAS.494.5091G} obtained using the spike statistics of the Ly$\alpha$ transmitted flux. This can be a potential new tool to constrain reionization parameters as it is usually not included in Bayesian parameter space explorations.\footnote{There exist other measurements of $T_0$ and $\gamma$ at redshifts of our interest \citep[see, e.g.,][]{2019ApJ...872...13W}. We choose the data of \citet{2020MNRAS.494.5091G} as our default because their estimates are claimed to be less sensitive to the uncertainties in the quasar continuum and the value of the photoionization rate. However, for the sake of completeness, we carry out a different analysis where we replace the \citet{2020MNRAS.494.5091G} data points with those from \citet{2019ApJ...872...13W}, see \appndx{app:alternates}.} It is worth noting that the $T-\Delta$ relation can show significant scatter during reionization and also after its completion, and hence cannot be described by a simple power-law. In fact, our modelling of the temperatures also shows scatter at redshifts $z \sim 6$, as discussed in Paper I. \citet{2020MNRAS.494.5091G} fit a power-law to the $T-\Delta$ relation in their radiative transfer simulations of patchy reionization even in the presence of the scatter and obtain $T_0$ and $\gamma$. They find that models with different $T_0$ and $\gamma$ predict different spike statistics and hence one can distinguish between a hot and a cold IGM using these parameters even in the presence of the scatter.

\end{enumerate}

\section{Likelihood and parameter space exploration} 
\label{sec:parms_stat}

\begin{table*}
\caption{Parameter constraints obtained from the MCMC-based analysis for the presently available data. The first six rows correspond to the free parameters of the model while the others are the derived parameters. The free parameters are assumed to have uniform priors in the range mentioned in the second column. The other numbers are showing the mean value with 1$\sigma$ errors for different parameters using different combinations of MCMC runs. The bestfit values are quoted within the brackets for the different parameters.}
\begin{threeparttable}
\begin{tabular}{cccccc}
\hline
Parameters & Prior & default & w/o-temp-data & grad-fb & clump-free\\
\hline
$\log(\zeta_0)$     & [$0,\infty$]     & $0.81^{+0.37}_{-0.32}~(0.92)$ & $0.85^{+0.41}_{-0.34}~(1.23)$ & $0.78^{+0.36}_{-0.31}~(0.96)$ & $0.79^{+0.36}_{-0.33}~ (0.79)$ \\ \\
$\alpha$ & [$-\infty,\infty$]  & $3.0^{+1.6}_{-2.1}~(2.3)$ & $3.1^{+1.7}_{-2.3}~(1.2)$ & $3.9^{+1.9}_{-2.3}~(2.2)$ & $3.0^{+1.6}_{-2.0}~(2.1)$ \\ \\
$T_{\mathrm{re}}$ ($10^4$~K)    & [$0.01, 10$]      & $1.85^{+0.23}_{-0.27}~(1.83)$ & $ < 3.32~(0.23)$ & $2.00^{+0.28}_{-0.35}~(1.93)$ & $1.86^{+0.25}_{-0.29}~(1.93)$ \\ \\
$f_{\mathrm{esc}}^0$   & [$0,1$]       & $0.26^{+0.06}_{-0.06}~(0.36)$ & $0.28^{+0.08}_{-0.09}~(0.36)$    & $0.30^{+0.08}_{-0.08}~(0.36)$ & $0.26^{+0.08}_{-0.07}~(0.30)$ \\ \\
$\beta$     &  [$-1,0$]         & $-0.38^{+0.07}_{-0.08}~(-0.44)$ & $-0.36^{+0.08}_{-0.09}~(-0.42)$  & $-0.34^{+0.08}_{-0.09}~(-0.44)$ & $-0.39^{+0.07}_{-0.09}~(-0.50)$ \\ \\
$C_{\mathrm{HII}}$\tnote{a} & [$0,7$] & $3$ & $3$ &$3$& $C_{\mathrm{HII}} < 4.0~(0.55)$ \\ \\
\hline
Derived Parameters & & & & & \\ \\
$\tau_e$& &$0.053^{+0.004}_{-0.006}~(0.053)$ & $0.054^{+0.005}_{-0.006}~(0.055)$ & $0.054^{+0.004}_{-0.006}~(0.054)$ & $0.053^{+0.005}_{-0.006}~(0.054)$ \\ \\
$\Delta z$& &$1.81^{+0.51}_{-0.67}~(1.87)$ & $1.80^{+0.52}_{-0.71}~(1.98)$ & $1.62^{+0.43}_{-0.59}~(1.95)$ & $1.81^{+0.50}_{-0.65}~(1.90)$ \\ \\
$z_{\mathrm{mid}}$ &  &$7.00^{+0.30}_{-0.40}~(7.04)$ & $7.12^{+0.36}_{-0.48}~(7.16)$ & $7.19^{+0.36}_{-0.44}~(7.14)$ & $7.04^{+0.32}_{-0.42}~(7.07)$ \\ \\
$z_{\mathrm{75}}$ & &$6.32^{+0.14}_{-0.17}~(6.34)$ & $6.44^{+0.22}_{-0.35}~(6.44)$ & $6.59^{+0.26}_{-0.35}~(6.41)$ & $6.36^{+0.16}_{-0.22}~(6.37)$ \\
\hline
\end{tabular}
\begin{tablenotes}
\item[a] $C_{\mathrm{HII}}$ is fixed to a value $3$ for the default, w/o-temp-data and grad-fb runs while it is kept a free parameter for the clump-free run.
\end{tablenotes}
\end{threeparttable}
\label{tab:param_cons}
\end{table*}

We use a Bayesian analysis to constrain the free parameters of our model. Our aim is to compute the conditional probability distribution or the posterior $\mathcal{P}(\lambda \vert \mathcal{D})$ of the model parameters $\lambda$ given the observed data sets $\mathcal{D}$ mentioned in the previous section. This can be computed using the Bayes theorem
\be\label{eq:bayes_eq}
 \mathcal{P}(\lambda\vert \mathcal{D})=\frac{\mathcal{L}(\mathcal{D} \vert \lambda) ~\pi(\lambda)}{\mathcal{P}(D)},
\ee
where $\mathcal{L}(\mathcal{D} \vert \lambda)$ is the conditional probability distribution of data given the parameters or the likelihood, $\pi(\lambda)$ is the prior and $\mathcal{P}(\mathcal{D})$ is the evidence (which can be treated as the normalization parameter and does not play any role in our analysis). The likelihood is assumed to be multi-dimensional gaussian
\bear
\label{eq:chisq_eq}
\mathcal{L}(\mathcal{D} \vert \lambda) 
&= \exp \left(-\frac{1}{2} \sum_{i}\left[\frac{\mathcal{D}_i-\mathcal{M}_i(\lambda)}{\sigma_i}\right]^2 \right)
\nline
&= \prod_i \exp \left(-\frac{1}{2} \left[\frac{\mathcal{D}_i-\mathcal{M}_i(\lambda)}{\sigma_i}\right]^2 \right),
\ear
where $\mathcal{D}_i$ are the measured values of the data points, $\mathcal{M}_i(\lambda)$ are the model predictions for the parameters $\lambda$ and $\sigma_i$ are the observational error bars on the data. The summation index $i$ runs over all data points used in the analysis. 

In some cases, we do have only a limit on the data point instead of a measurement. In such cases, the likelihood is given by \citep{2020MNRAS.493.4728G,2021MNRAS.500.5322G}

\be\label{eq:chi_lim}
\mathcal{L}(\mathcal{D} \vert \lambda)  = \prod_i \frac{1}{2}\left[1+\mathrm{erf}\left(\pm\frac{\mathcal{M}_i(\mathcal{\lambda})-\mathcal{D}_i}{\sqrt{2}~\sigma_i}\right)\right],
\ee
where the $\pm$ sign inside the error function represents the lower and upper limits respectively. 

We sample the posterior distribution using the Monte Carlo Markov Chain (MCMC) method, more specifically, the Metropolis-Hastings algorithm \citep{1953JChPh..21.1087M}. We make use of the publicly available package \texttt{cobaya} \citep{2021JCAP...05..057T}\footnote{\url{https://cobaya.readthedocs.io/en/latest/}} to run the MCMC chains.  The samples are drawn using 12 parallel chains \citep{2002PhRvD..66j3511L,2013PhRvD..87j3529L}. The chains are assumed to converge when the Gelman-Rubin $R - 1$ statistic \citep{1992StaSc...7..457G} becomes less than a threshold $0.01$. This typically needs around $10^5$ steps for our model and completes in about 1-2 days. We discard first few steps in the chains, equal to $30\%$ of the total samples, as `burn-in' and work with only the rest.

\section{Results}
\label{sec:results}

In this section, we present our constraints on the reionization history and the free parameters of the model. Our default likelihood analysis involves five parameters of the model and utilize all the observational data sets discussed in \secn{sec:obs_cons}. To understand the importance and physical implications of the results obtained with the default run, we also present a few variants. Let us first discuss these different MCMC runs:

\begin{itemize}

\item \textbf{default:} As mentioned above, we vary the five free parameters $ \{\log \zeta_0,\alpha,T_{\mathrm{re}},f_{\mathrm{esc}}^0,\beta\}$. We set $C_{\mathrm{HII}} = 3$ and use the step feedback prescription. We use all the five data sets outlined in \secn{sec:obs_cons} for this run.

\item \textbf{w/o-temp-data:} In this run, we \emph{neglect} the $T_0$ and $\gamma$ measurements at $z \sim 5.5-6$. The aim of this run is to understand the importance of the temperature data on reionization history constraints. 

\item \textbf{grad-fb:} To study the effect of the feedback prescription on the parameter constraints, we employ a MCMC run where the feedback is taken to be gradual (instead of step as in the default model).

\item \textbf{clump-free:} Finally, we use an analysis where the clumping factor $C_{\mathrm{HII}}$ is kept as a free parameter. Hence we vary six parameters in this run.

\end{itemize}

The parameter priors and the constraints on them obtained from different MCMC runs are summarized in \tab{tab:param_cons}. We assume flat priors for all the free parameters. The priors on $\log \zeta_0$, $\alpha$ and $f_{\mathrm{esc}}^0$ span the whole possible range. The prior on the reionization temperature $T_{\mathrm{re}}$ is assumed to be in the range $[100, 10^5]$~K which is considerably wider than the typical values found in radiative transfer simulations \citep{2012MNRAS.426.1349M,2018MNRAS.480.2628F,2019ApJ...874..154D}. The prior range of $\beta$ too is kept sufficiently wider than the constraints we obtain. The clumping factor $C_{\mathrm{HII}}$ is free in the clump-free run where we allow it to be within $[0,7]$ while it is fixed to $3$ for the other runs.

\begin{figure*}
    \includegraphics[width=0.99\textwidth]{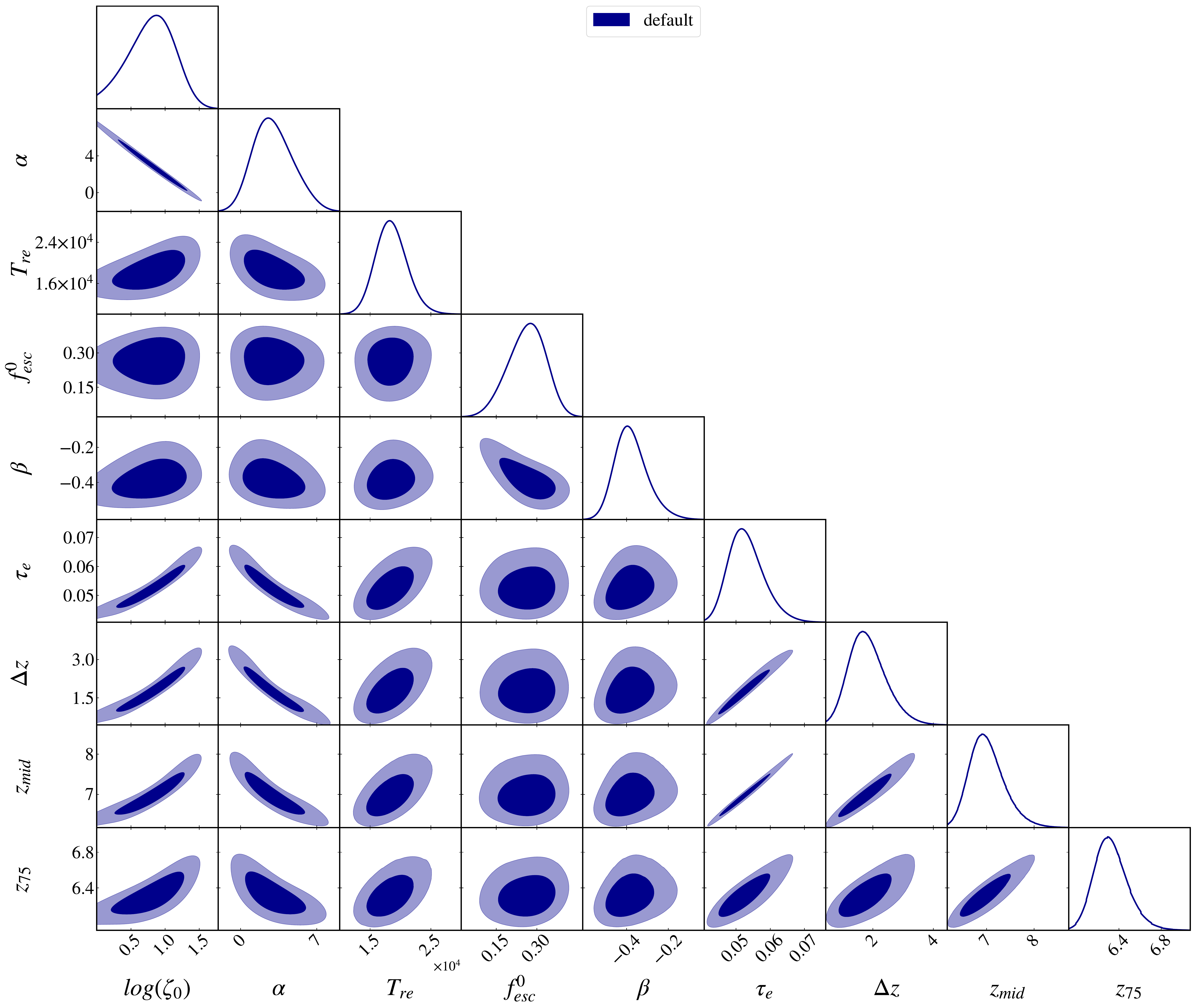}
    \caption{Posterior distributions of the five free and four derived parameters of our model for the default case. In the diagonal panels, we have shown the plots of marginalized one-dimensional posterior probabilities. The off-diagonal panels show the marginalized two-dimensional joint probability distributions. The contours represent the 68\% and 95\% confidence intervals.}
    \label{fig:corner_default}
\end{figure*}

\begin{figure}
    \includegraphics[width=0.99\columnwidth]{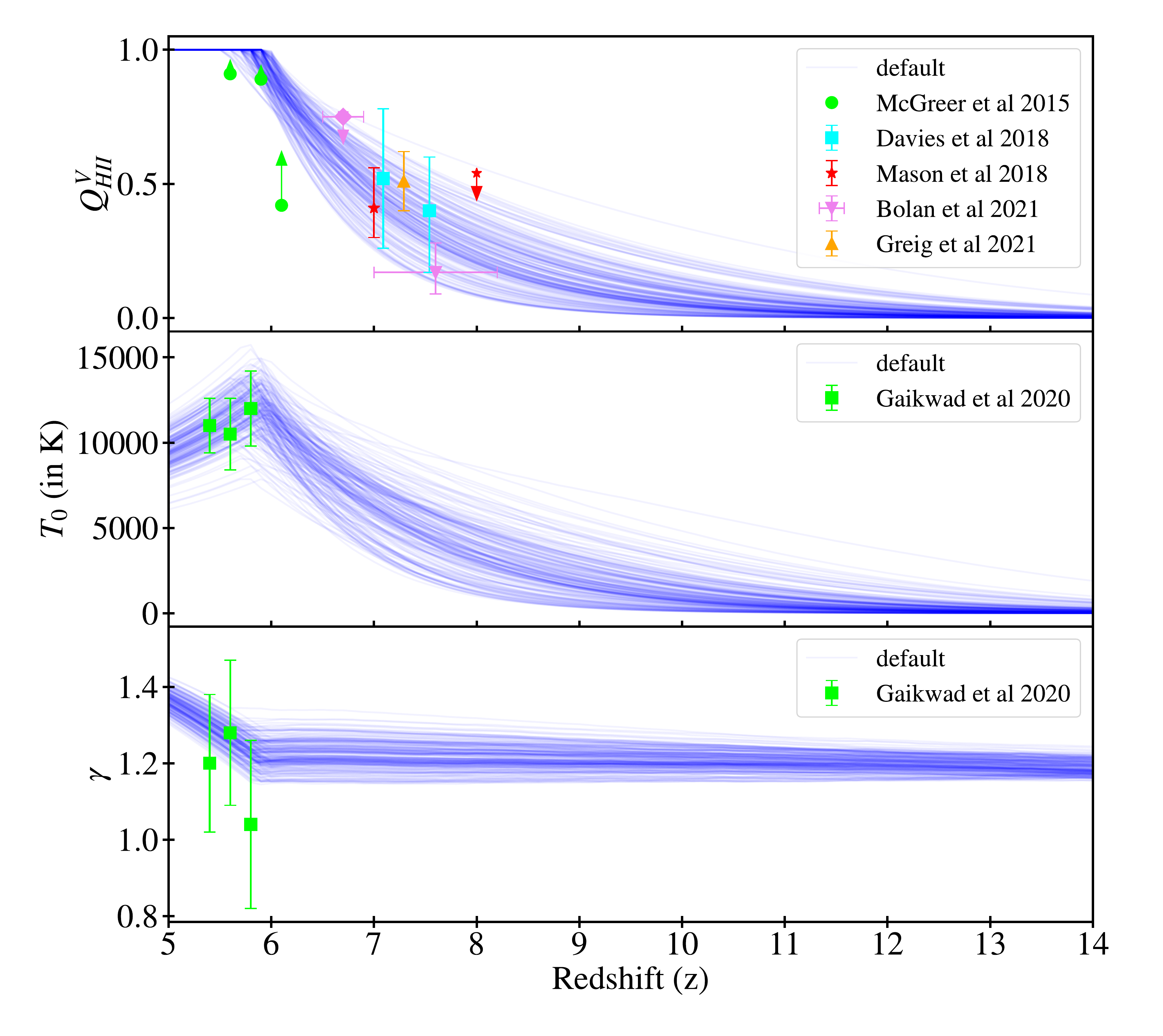}
    \caption{The reionization (top panel) and thermal (middle and bottom panels) histories for 200 random samples drawn from the MCMC chains for the default run. The data points show the different observational constraints with the source mentioned in the legend. Note that only the \citet{2015MNRAS.447..499M} data is used in the MCMC analysis and the other points are shown only for comparison with our results.}
    \label{fig:hist_default}
\end{figure}

\begin{figure}
    \includegraphics[width=0.99\columnwidth]{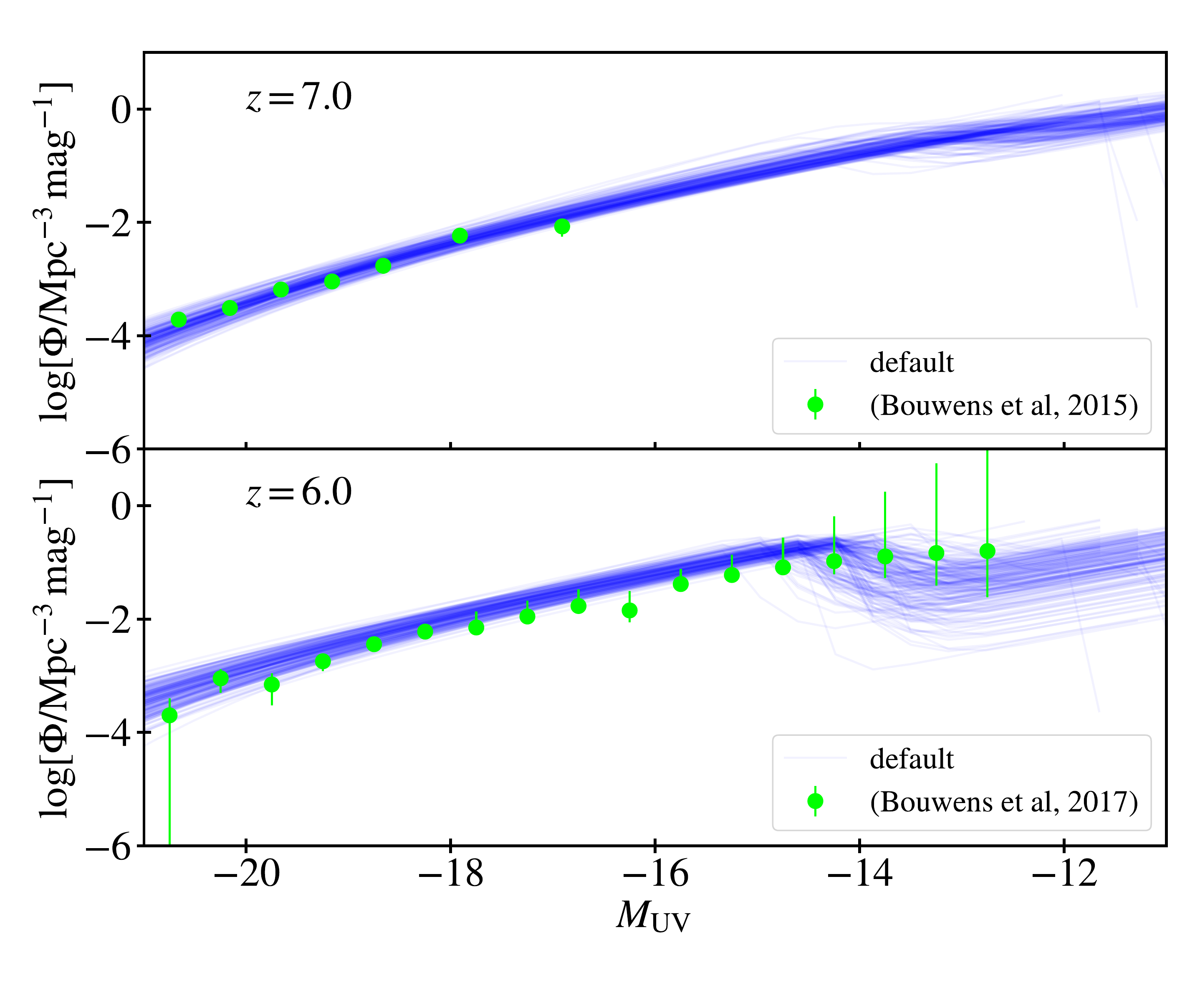}
    \caption{The galaxy UV luminosity functions at $z = 7$ (top panel) and $6$ (bottom panel) for 200 random samples drawn from the MCMC chains for the default run. The data points show the observational constraints of \citet{2015ApJ...803...34B,2017ApJ...843..129B}.}
    \label{fig:uvlf_default}
    \end{figure}

\begin{figure}
    \includegraphics[width=0.99\columnwidth]{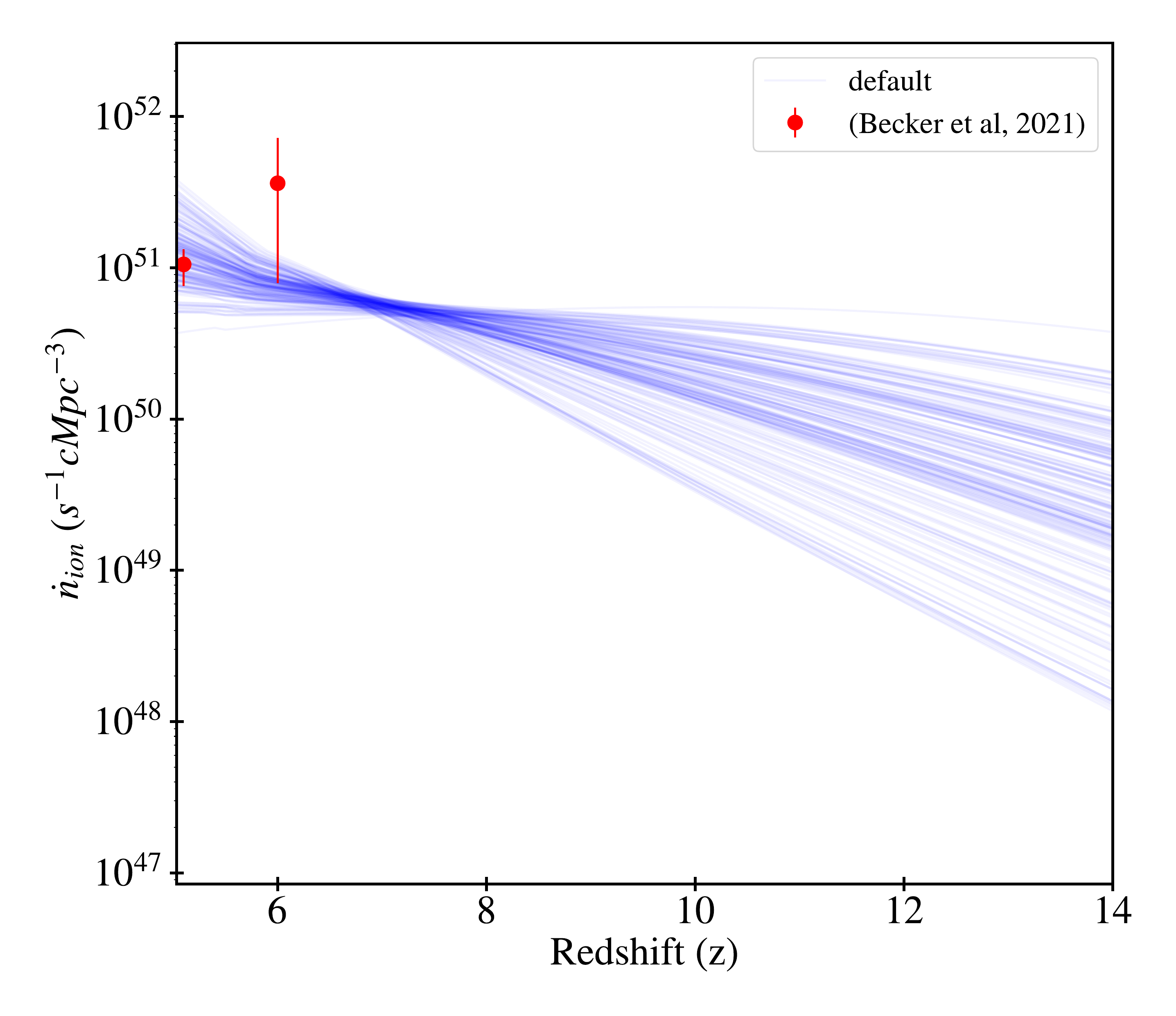}
    \caption{The evolution of the ionizing emmissivities for 200 random samples drawn from the MCMC chains for the default run. The data points are indirect estimates of the emmissivities obtained from the measurements of the HI photoionization rates and the mean free paths of ionizing photons.}
    \label{fig:emissivity_default_becker}
    \end{figure}

\begin{figure}
    \includegraphics[width=0.99\columnwidth]{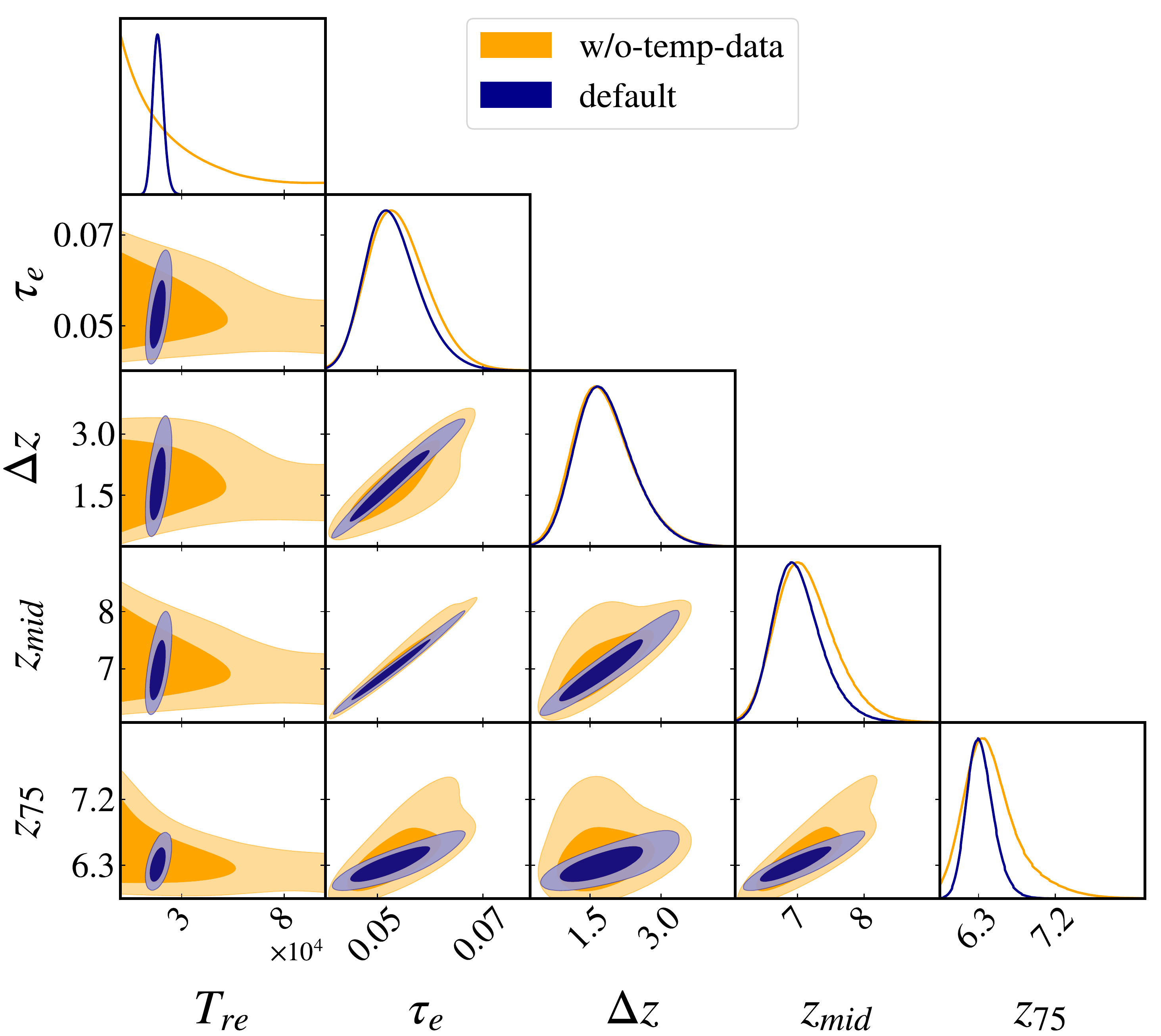}
    \caption{Posterior distributions of a subset of free and derived parameters of the model for two cases: default (blue contours, identical to \fig{fig:corner_default}) and w/o-temp-data (orange contours). The contours represent the 68\% and 95\% confidence intervals.}
    \label{fig:corner_default_wotemp}
\end{figure}

\begin{figure}
    \includegraphics[width=0.99\columnwidth]{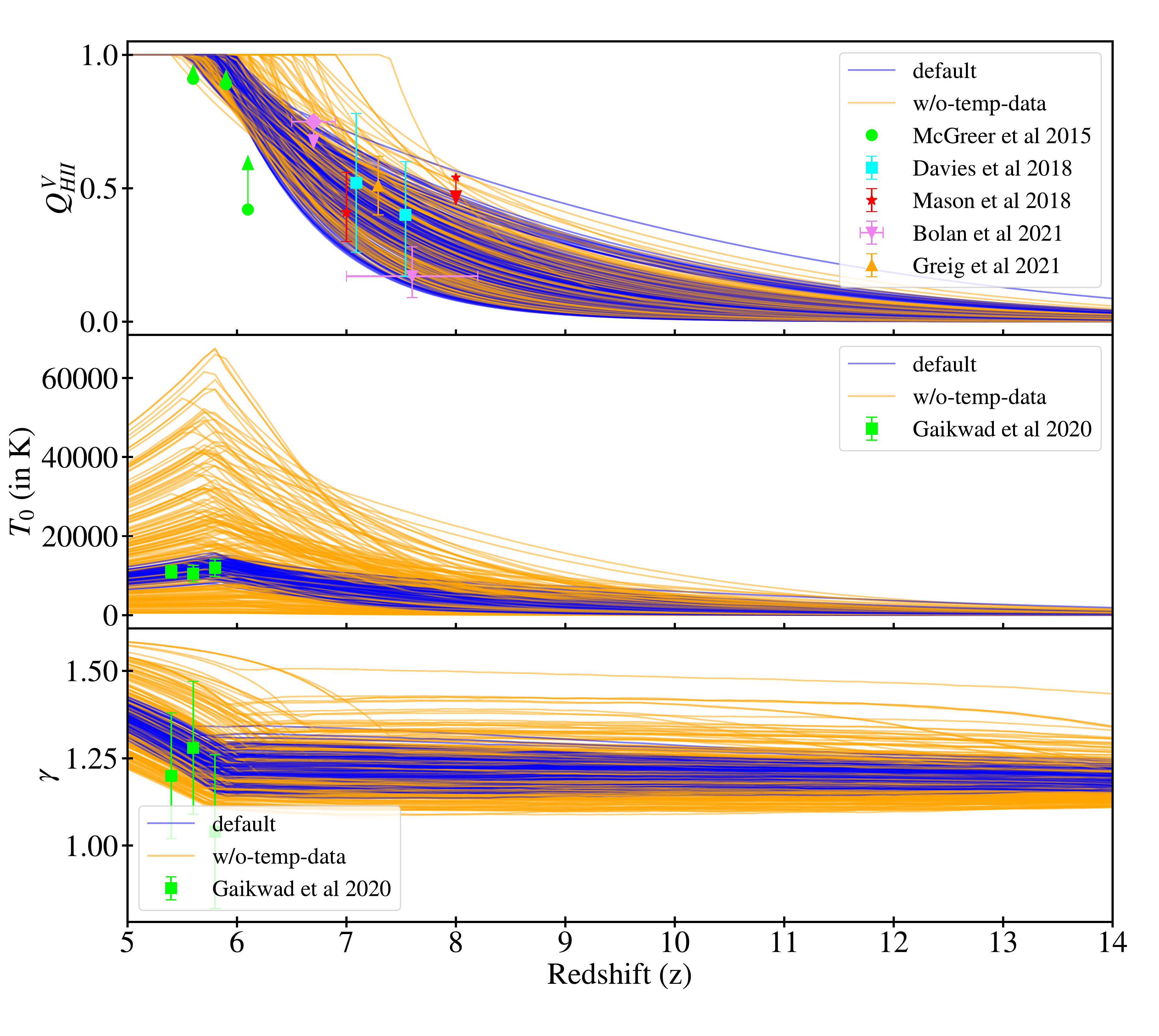}
    \caption{Same as \fig{fig:hist_default} except that the random samples drawn from the MCMC chains are shown both for the default (blue) and the w/o-temp-data (orange) cases.}
    \label{fig:hist_default_wotemp}
\end{figure}

\begin{figure}
    \includegraphics[width=0.99\columnwidth]{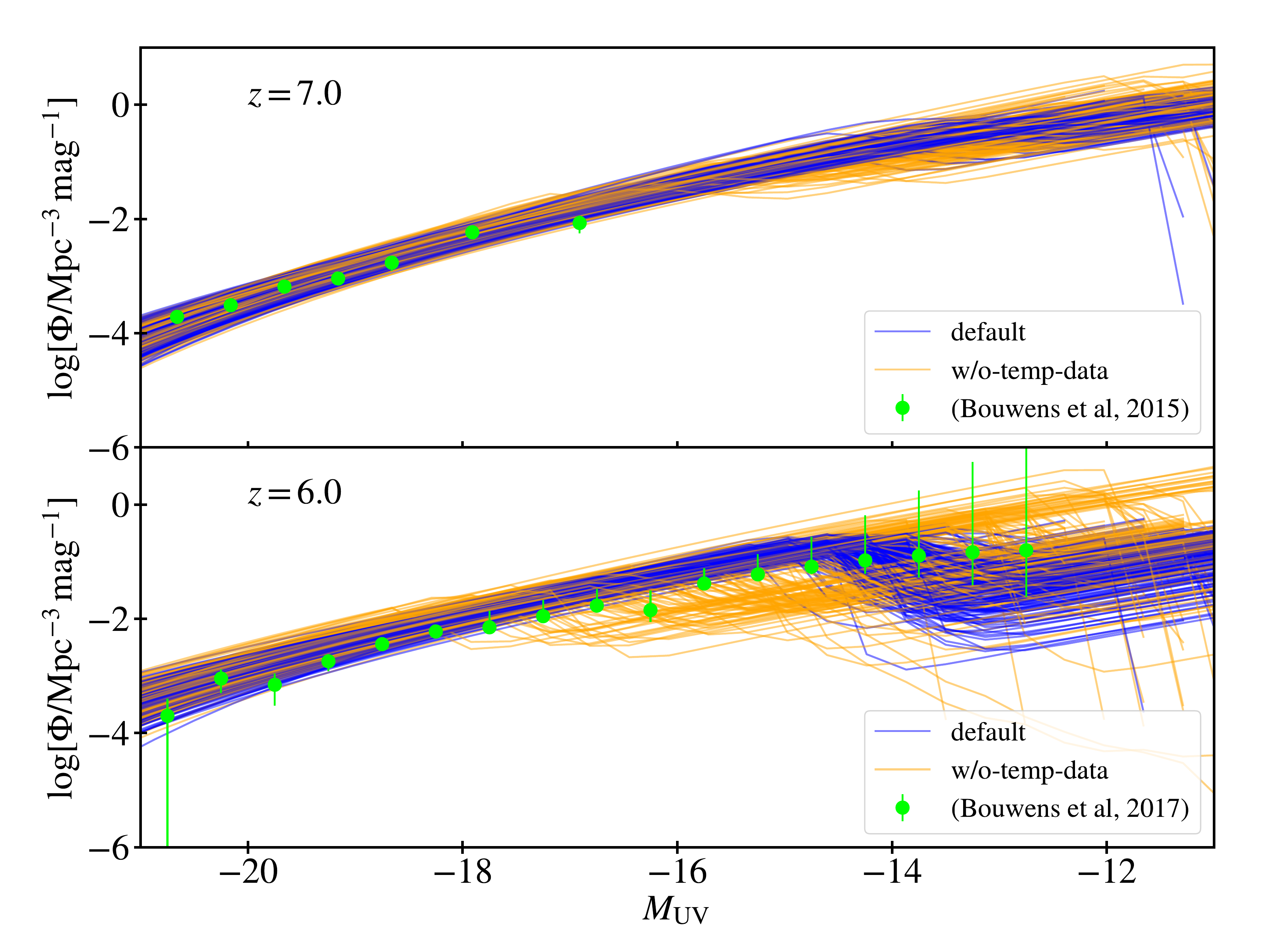}
    \caption{Same as \fig{fig:uvlf_default} except that the random samples drawn from the MCMC chains are shown both for the default (blue) and the w/o-temp-data (orange) cases.}
    \label{fig:uvlf_default_wotemp}
\end{figure}

\begin{figure}
    \includegraphics[width=0.99\columnwidth]{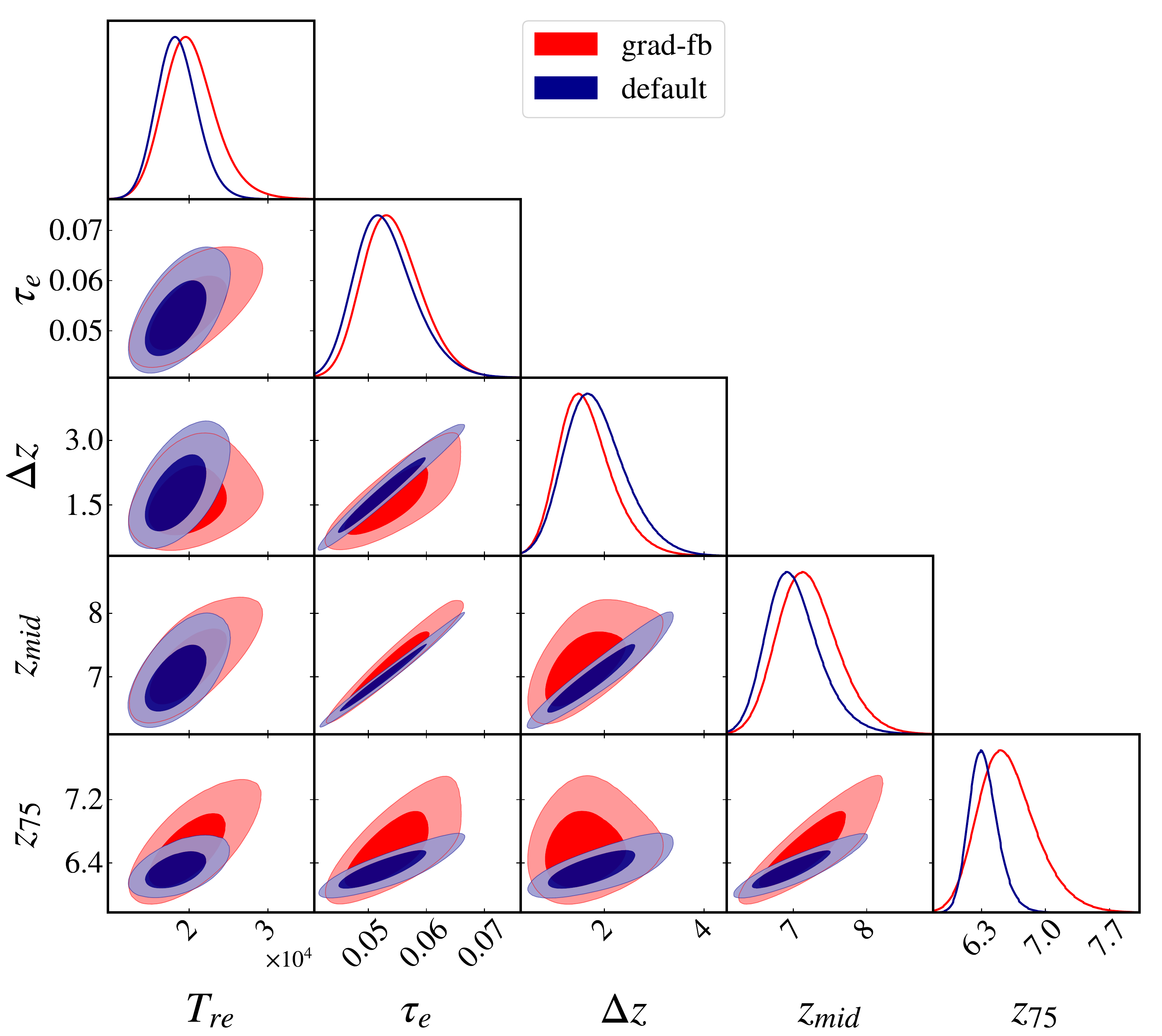}
    \caption{Posterior distributions of a subset of free and derived parameters of the model for two cases: default (blue contours, identical to \fig{fig:corner_default}) and grad-fb (red contours). The contours represent the 68\% and 95\% confidence intervals.}
    \label{fig:corner_default_grad_fb}
\end{figure}

\begin{figure}
    \includegraphics[width=0.99\columnwidth]{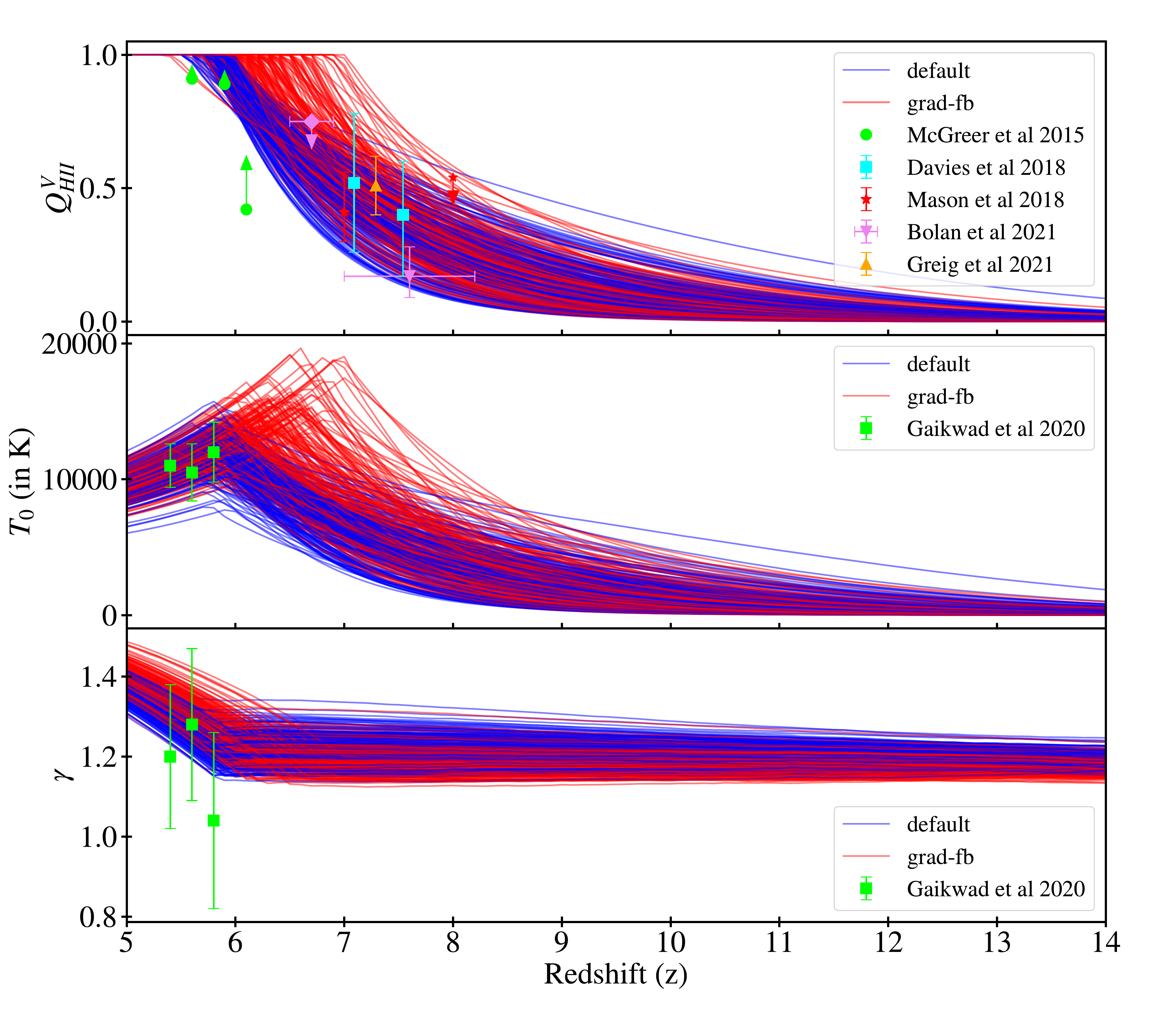}
    \caption{Same as \fig{fig:hist_default} except that the random samples drawn from the MCMC chains are shown both for the default (blue) and the grad-fb (red) cases.}
    \label{fig:hist_default_grad_fb}
\end{figure}

\begin{figure}
    \includegraphics[width=0.99\columnwidth]{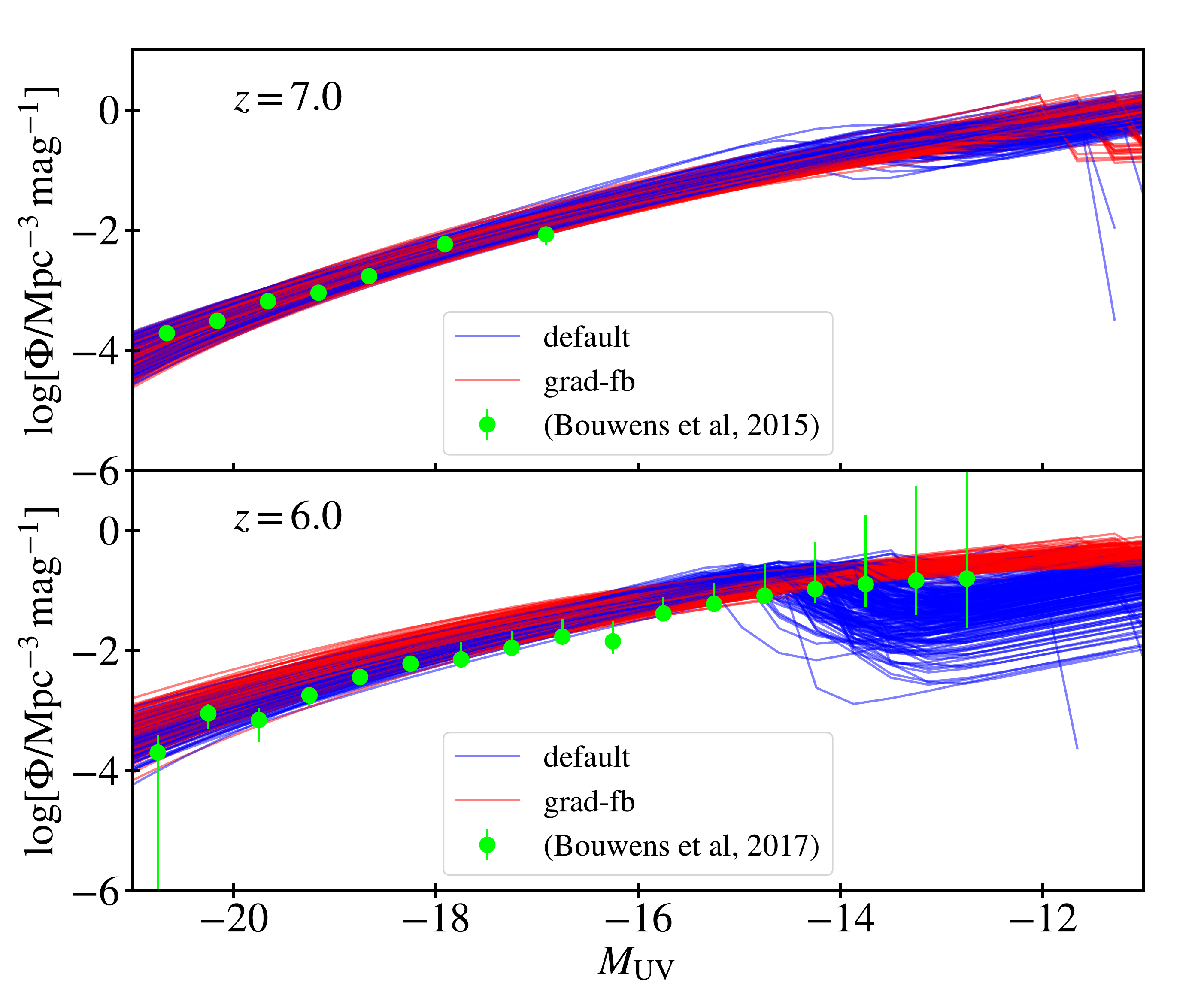}
    \caption{Same as \fig{fig:uvlf_default} except that the random samples drawn from the MCMC chains are shown both for the default (blue) and the grad-fb (red) cases.}
    \label{fig:uvlf_default_grad_fb}
\end{figure}
    
Let us begin the discussion with the results obtained from the default run. The posterior distributions of the free and derived parameters are shown in \fig{fig:corner_default}. The corresponding constraints on the parameters can be found in \tab{tab:param_cons}. We find that the individual constraints on the parameters that characterize the ionization efficiency, i.e., $\zeta_0$ and $\alpha$, are not that stringent. The range in $68\%$ confidence limits for $\log \zeta_0$ is $\sim 0.7$~dex while the slope $\alpha$ has a range $\sim 4$. The models mildly favour $\alpha > 0$ implying that the ionization efficiency should be higher at lower redshifts. There is a strong anti-correlation between $\log \zeta_0$ and $\alpha$ as can be seen from the same figure. From the form of $\zeta(z)$ in \eqn{eq:zeta_eq}, it is clear that, for a fixed $\zeta_0$, lower values of $\alpha$ leads to less efficient ionizing sources at $z < 9$. In order to match the observations, this decrease needs to be compensated by increasing the normalization $\zeta_0$. One can use the exact form of the degeneracy between $\log \zeta_0$ and $\alpha$ to show that the efficiency $\zeta(z)$ is rather tightly constrained at $z \sim 5.6$.

Even though the constraints on $\zeta_0$ and $\alpha$ are somewhat weak, the allowed range in the resulting reionization histories are rather tight. This can be verified from the constraints on $\tau_e$, $\Delta z$, $z_{\mathrm{mid}}$ and $z_{75}$. The typical behaviour of the favoured reionization histories can be understood from the top panel of \fig{fig:hist_default} where we plot the evolution of the ionized fraction of 200 random samples from the MCMC chains. The end of reionization, in particular, is quite tightly constrained around $z \lesssim 6$ which, as we will discuss later, is a consequence of including the temperature measurements at $z \sim 5.5$ in the analysis. For comparison, we also plot constraints on the ionized fraction obtained from studies not considered in our analysis, e.g., those from the damping wing signatures in the quasar spectra \citep{2018ApJ...864..142D,2021arXiv211204091G} and from the Ly$\alpha$ emission detected in Lyman-break galaxies \citep{2018ApJ...856....2M,2021arXiv211114912B}. As is clear from the figure, these constraints are consistent with those obtained from our analysis.

The inclusion of the UVLF observations allows us to put constraints on the escape fraction. From the constraints on $f_{\mathrm{esc}}^0$, we find that we require escape fractions $\sim 0.26$ for $M_h = 10^9 \Msun$ haloes at $z = 6 - 7$. We also find that the slope $\beta$ is negative with a mean value $-0.38$ and $\beta = 0$ is ruled out at high significance $\gtrsim 4\sigma$. Because of the assumptions involved, the constraint on $\beta$ implies that the star-forming efficiency $f_* \propto M_h^{0.38}$ increases with halo mass. This is consistent with other works where one required a power-law index $\sim 0.5$ to match the UVLF observations \citep{2019MNRAS.484..933P,2021MNRAS.506.2390Q}. Since we have assumed $\zeta \propto f_* f_{\mathrm{esc}}$ to be independent of $M_h$, the constraint on $\beta$ also indicates that the escape of ionizing photons from the high-mass haloes is less efficient.  The $M_h$-dependence of $f_{\mathrm{esc}}$ implied by our analysis is similar to what is found in different hydrodynamic simulations of galaxy formation \citep{2015MNRAS.451.2544P,2016ApJ...833...84X,2020MNRAS.498.2001M}. The mass-dependence agrees also with other parameter statistics analysis using semi-numerical models \citep{2019MNRAS.484..933P,2021MNRAS.506.2390Q}. There exist observational estimates of $f_{\mathrm{esc}}$ at relatively lower redshifts \citep{2018A&A...616A..30C,2020A&A...639A..85G,2021MNRAS.508.4443M} which indicate similar values. Similar to \fig{fig:hist_default}, we show in \fig{fig:uvlf_default} the UVLF for the same 200 random samples, along with the data points of \citet{2015ApJ...803...34B,2017ApJ...843..129B} used in the MCMC analysis.

The parameter related to the thermal evolution, $T_{\mathrm{re}}$ is constrained to values $\sim 2 \times 10^4$~K (see \tab{tab:param_cons}). It is correlated with the reionization parameters, but perhaps the most interesting ones are the correlations with the derived parameters $\tau_e$, $\Delta z$, $z_{\mathrm{mid}}$ and $z_{75}$. The positive correlations between $T_{\mathrm{re}}$ and these parameters essentially imply that higher values of $T_{\mathrm{re}}$ leads to relatively early reionization histories. Since the IGM starts cooling once reionization is complete, early reionization histories would lead to more cooling of the gas at $z \sim 5.5$ (redshifts where measurements of the thermal parameters exist). Hence to match the observations, one would need a higher reionization temperature for such models. This can also be understood from the thermal evolution of 200 random samples drawn from the MCMC chains, as is shown in the middle and bottom panels of \fig{fig:hist_default}. At this stage, one may wonder that if $T_{\mathrm{re}}$ and the redshift corresponding to the end of reionization are so degenerate, how do we obtain reasonably strong constraints on them individually. This has partly to do with the constraints on $\tau_e$ which rule out very early reionization and hence very high $T_{\mathrm{re}}$. Very high values of $T_{\mathrm{re}}$ are ruled out also by the UVLF observations at $z = 6$. According to our model, since high $T_{\mathrm{re}}$ would lead to more severe radiative feedback, it would start affecting the faint end of the luminosity function and hence can be ruled out by the data. Note, however, that the constraints on $T_{\mathrm{re}}$ depend on the feedback model used, as will become clear when we discuss some of the alternate cases later.) This can be seen in \fig{fig:uvlf_default} (bottom panel) where feedback effects start to become prominent at $M_{\mathrm{UV}} \gtrsim -14$ and can be distinguished by the observational data points. This shows the interplay between different data sets in providing constraints on the parameters and hence on the reionization and thermal histories.

Before moving on, let us compute the emissivity of ionizing photons $\dot{n}_{\mathrm{ion}}$ as implied by the parameter constraints. We show the evolution of $\dot{n}_{\mathrm{ion}}$ for the 200 randomly chosen models in \fig{fig:emissivity_default_becker}. The parameter degeneracies act in a way such that the emissivity is relatively tightly constrained at $z \sim 7$. The constraints are rather poor at very high redshifts and also weaker at $z \lesssim 7$. To understand whether these emissivities are sensible, we compare with observational estimates of the same. There are no direct measurements of $\dot{n}_{\mathrm{ion}}$, however, one can provide an estimate from those of the HI photoionization rate $\Gamma_{\mathrm{HI}}$ and the mean free path $\lambda_{\mathrm{mfp}}$ of ionizing photons (see Paper I for more details). Both these have been estimated at $z \sim 5-6$ using the quasar absorption spectra and hydrodynamical simulations \citep{2011MNRAS.412.1926W,2013MNRAS.436.1023B,2021MNRAS.508.1853B,2021ApJ...917L..37C}. We use the data of \citet{2021MNRAS.508.1853B} to estimate $\dot{n}_{\mathrm{ion}}$ along with the observational error bars. These are shown in \fig{fig:emissivity_default_becker} by red points. It is clear that the emissivities allowed by our analysis are consistent with the observational estimates allowing for the error bars. Interestingly, the central value of $\dot{n}_{\mathrm{ion}}$ at $z = 6$ is larger than what is allowed by the MCMC constraints. On the other hand, the corresponding data at $z = 5$ has much smaller error bar than the spread of the models. Hence including the observational constraints of $\Gamma_{\mathrm{HI}}$ and $\lambda_{\mathrm{mfp}}$ may lead to tighter constraints on the reionization histories which we plan to take up in the near future.

\subsection{Consequence of including the temperature measurements}

Let us now understand the importance of including the measurements of $T_0$ and $\gamma$ at $z \sim 5.5$ in the analysis. These measurements are usually not accounted for while studying parameter constraints related to reionization. To appreciate their importance, we run a MCMC chain, named w/o-temp-data, where we do \emph{not} include the contribution of these measurements in the likelihood. The posterior distributions of a selected set of parameters are shown in \fig{fig:corner_default_wotemp}, both for the default run (blue contours) and for the w/o-temp-data (orange contours). The 200 random samples drawn from the MCMC chains for each of the runs are shown in \fig{fig:hist_default_wotemp} and \fig{fig:uvlf_default_wotemp}. The corresponding parameter constraints can be found in \tab{tab:param_cons}.

The obvious, and possibly expected, difference between the two runs is that the constraint on $T_{\mathrm{re}}$ is significantly weakened once the temperature data is not included in the analysis. In fact, we have only an upper limit $T_{\mathrm{re}} < 3.32 \times 10^4$~K ($68\%$ confidence limit) on the reionization temperature. This limit arises because higher temperatures lead to so severe a feedback that the UVLF starts to get affected even at relatively brighter magnitudes (see, e.g., the orange lines in \fig{fig:uvlf_default_wotemp}). 

One important consequence of higher allowed values of $T_{\mathrm{re}}$ in the w/o-temp-data run is that histories where reionization completes early become allowed by the data. As we can see from the random samples in \fig{fig:hist_default_wotemp}, several reionization histories, where reionization completed at $6 \lesssim z \lesssim 7$, are not favoured in the default run, however, they are allowed when the temperature data is ignored. This is also evident from the joint posterior distributions between $T_{\mathrm{re}}$ and the parameters $\tau_e$, $\Delta z$, $z_{\mathrm{mid}}$, $z_{75}$ shown in \fig{fig:corner_default_wotemp}. The analysis in this section thus leads to an important conclusion that incorporating the temperature data in the analysis disfavours models with early reionization histories.

\subsection{Effect of the feedback prescription}

We next check the effect of the feedback prescription on the parameter constraints. Our default run assumes the feedback to be step-like. As a variant of the default run, we use a prescription where feedback is gradual, the run named grad-fb (see \secn{sec:brief_model} for more details on the prescription). In \fig{fig:corner_default_grad_fb}, we show the comparisons of the posteriors for a subset of parameters between the default (\emph{blue}) and grad-fb (\emph{red}) runs. The parameters chosen for demonstration are $T_{\mathrm{re}}$ and the derived parameters characterizing the reionization history. It is clear that the constraints on the parameters are somewhat weakened when the feedback is gradual. In particular, the grad-fb run allows for larger $T_{\mathrm{re}}$ values than the default one. Because of the degeneracies between $T_{\mathrm{re}}$ and the other parameters, this immediately implies that relatively earlier completion of reionization is allowed when the feedback is gradual. This can also be seen from \fig{fig:hist_default_grad_fb} where we show plots of 200 random samples from the MCMC chains. The red curves which correspond to the grad-fb case show early completion of reionization as compared to the blue ones for the default run. We show the UVLF for these samples in \fig{fig:uvlf_default_grad_fb}. At the faint end, the effect of radiative feedback is much less severe for the grad-fb models (red curves) than that for the default models (blue curves) with step feedback.

To understand the reason for this difference between the default and grad-fb runs, note that, for the same $T_{\mathrm{re}}$, the gradual feedback prescription affects the low-mass haloes less severely. Hence one can allow higher values of $T_{\mathrm{re}}$ without affecting the faint end of the UVLF significantly, see the UVLF trends in \fig{fig:uvlf_default_grad_fb}. Now, the survival of star-formation in the low-mass haloes for longer times in the grad-fb model allows for earlier reionization. Since the IGM cools once reionization is complete, these early reionization histories would require higher $T_{\mathrm{re}}$ to match the temperature measurements at $z \sim 5.5$. The chain of reasoning thus leads to the understanding that the data sets allow larger $T_{\mathrm{re}}$ values and early completion of reionization when the feedback is gradual.

From \tab{tab:param_cons}, we also find that the reionization duration $\Delta z$ is slightly smaller in the grad-fb case compared to the default. When the feedback affects the low-mass haloes less severely, it leads to faster evolution of reionization. This too is expected as the feedback acts as a regulation mechanism which works against any rapid growth of ionization fraction. Clearly, more severe the feedback, more efficient would be the regulation.

Since the nature of reionization constraints depends in the feedback prescription, it becomes important to find a way to constrain it observationally. As far as the scope of this paper goes, the best prospect is to have better data at the faint end of the UVLF, both at $z = 6$ and $7$. A possibility which is likely to open up in the near future is the JWST data \citep{2015ApJ...813...21M,2020MNRAS.491.3891P,2020MNRAS.499.5702B}, although their effectiveness in probing galaxies affected by feedback is still debated \citep{2018MNRAS.474.2352C,2020MNRAS.492.5167V}.

\subsection{Clumping factor as a free parameter}

We have also checked the parameter posteriors keeping the clumping factor $C_{\mathrm{HII}}$ as a free parameter. The run is named clump-free and the corresponding parameter constraints are given in \tab{tab:param_cons}. We assumed a flat prior in the range $[0,7]$ for $C_{\mathrm{HII}}$. We find a weak $1\sigma$ limit of $<4.0$ and, in fact, there are essentially no constraints in this prior range at the $2\sigma$ confidence level. What is interesting is that the constraints on the other parameters are consistent with the default run where $C_{\mathrm{HII}}$ was fixed to $3$. Thus we conclude that the current observations cannot be used to put limits on $C_{\mathrm{HII}}$. At the same time, the lack of the knowledge on the clumping factor should not affect constraints on the other parameters or the reionization history.

\subsection{Comparison with other constraints obtained using SCRIPT}

We end this section with a discussion on how the results of this paper compare with similar constraints obtained with different version of our code. There are two other versions of \texttt{SCRIPT} which has been used for obtaining constraints on the reionization history.

In one of the earlier versions, which does \emph{not} include the inhomogeneous recombinations, the thermal evolution and the radiative feedback, we have compared the model predictions with the CMB data (not only $\tau_e$ but also the kinetic Sunyaev-Zeldovich signal from patchy reionization at small angular scales). The effect of radiative feedback was indirectly accounted for by taking a $z$-dependent $M_{\mathrm{min}}$ and leaving it as a free parameter. In that work, we found that the reionization duration $\Delta z \sim 1.3$ \citep{2021MNRAS.501L...7C}, which is smaller than the mean values but consistent with the $1\sigma$ limits obtained in this work. It is thus clear that the presence of recombinations and feedback makes the reionization rather gradual and hence leads to a duration that is longer than when these effects are not included.

In another version of \texttt{SCRIPT}, we included a few additional modules to compute the Ly$\alpha$ opacity at $z \sim 5 - 6$ so that the results can be compared with quasar absorption data \citep{2021MNRAS.501.5782C}. The results obtained indicated that the reionization cannot complete before $z \sim 5.6$. Such late completion of reionization is consistent with the constraints from the default run used in this work. On the other hand, the grad-fb run seems to allow for histories where the reionization completes early. It would then be interesting to extend the present code, include calculations of the Ly$\alpha$ opacity and verify whether the constraints for the grad-fb analysis get revised.

\section{Summary \& Conclusions}
\label{sec:conc}

In the present work, we have used a semi-numerical code, \texttt{SCRIPT} (see Paper I), for modelling the reionization and thermal histories of the universe to explore the prospects of constraining reionization parameters. Our models include the effects of inhomogeneous recombinations and radiative feedback which are important for a realistic modelling of the high-redshift universe. The default version of the model has five free parameters, i.e., two parameters $\log \zeta_0, \alpha$ related to the ionization efficiency of the galaxies, the reionization temperature $T_{\mathrm{re}}$, and two parameters $f_{\mathrm{esc}}^0, \beta$ for the escape fraction of the ionizing photons. In addition, one needs to specify the feedback prescription and also the value of the globally-averaged clumping factor $C_{\mathrm{HII}}$.
 
We have used different existing observations at multiple wavelength bands. These include model-independent estimates of ionization fraction at $z \lesssim 6$ using dark pixel statistics \citep{2015MNRAS.447..499M}, the Lyman-break UV luminosity functions at $z \sim 6 - 7$ \citep{2015ApJ...803...34B,2017ApJ...843..129B} and the CMB scattering optical depth $\tau_e$ \citep{2020A&A...641A...6P}. In addition to these, we also utilized the temperature measurements of the low-density IGM at $z \sim 5.5$ obtained through a combination of Ly$\alpha$ absorption spectra and hydrodynamical simulations \citep{2020MNRAS.494.5091G}, a probe not that commonly used while carrying out parameter space exploration for reionization. 

The comparison between the model predictions and the observational data has been carried out using a MCMC-based Bayesian analysis. This leads to constraints on the free parameters of the model and consequently on the reionization and thermal history. The main findings of our analysis are:

\begin{itemize}

\item In general, the individual parameters $\zeta_0$ and $\alpha$, characterizing the evolution of the ionization efficiency $\zeta(z)$ are not well-constrained by the observations used in this work. However, the data indicates a strong degeneracy between these two parameters. This leads to a rather strong constraint on the ionizing emmissivity at $z \sim 7$.

\item Our Bayesian analysis is able to provide reasonably strong constraints on the allowed reionization history. For example, our default MCMC run (with step feedback and clumping factor $C_{\mathrm{HII}} = 3$) favours late reionization histories, completing at $z\lesssim 6$. We also derive constraints on reionization duration $\Delta z=1.81^{+0.51}_{-0.67}$ and the midpoint $z_{\mathrm{mid}}=7.0^{+0.30}_{-0.40}$ using the analysis.

\item We find that the posterior distribution of $T_{\mathrm{re}}$ is strongly correlated with parameters characterizing the reionization history. Histories where reionization completes early require higher $T_{\mathrm{re}}$ to match the temperature data at $z \sim 5.5$. In fact, including the temperature measurements in the analysis disfavours completion of reionization at $z \gtrsim 6$ as these would lead to too cool an IGM at $z \sim 6$. Any possible workaround to this by increasing the $T_{\mathrm{re}}$ is ruled out by the faint-end of the galaxy UVLF observations. Our constraint on $T_{\mathrm{re}}$ is  $\left(1.85^{+0.23}_{-0.27}\right) \times 10^4$~K for the default case, consistent with the values $\sim 2\times 10^4 $K found in radiative transfer simulations \citep{2019ApJ...874..154D}.

\item Using the observations of the UVLF at $z = 6$ and $7$, the escape fraction of ionizing photons from galaxies is found to be $\sim 0.26$ for $10^9 \Msun$ haloes and it decreases with increasing halo mass. The value we find is similar to what is found in other semi-numerical studies that attempt to match the observations \citep{2013MNRAS.428L...1M,2016MNRAS.457.4051K,2021MNRAS.507.2405C}, and also the $M_h$-dependence is qualitatively similar to that found in simulations \citep{2015MNRAS.451.2544P,2016ApJ...833...84X,2020MNRAS.498.2001M}. Our analysis also implies that the star-forming efficiency $f_*$ must increase with halo mass, which too is consistent with semi-numerical studies that match the UVLF \citep{2019MNRAS.484..933P,2021MNRAS.506.2390Q}.

\item We find that the parameter constraints and the constraints on the reionization history depends on the exact feedback prescription used. In general, prescriptions leading to less severe feedback (e.g., those which affect the gas in the small mass haloes in a gradual manner) allow for relatively early completion of reionization. In comparison, the value of the clumping factor has much less influence on the constraints.

\end{itemize}

While the present version of the code, \texttt{SCRIPT}, is useful for comparing with a wide variety of data sets, it is still not complete. A slightly different version of the code (which does not include the thermal evolution) was used for constraining the reionization history with only CMB data, i.e., the $\tau_e$ constraints and the kinetic Sunyaev-Zeldovich (kSZ) signal from patchy reionization at small angular scales \citep{2021MNRAS.501L...7C}. A natural question then arises: why cannot we include the kSZ signal in the present analysis and see if the constraints improve? The main difficulty in doing this at present is technical: the small-scale kSZ signal requires the simulations to be run at higher resolution than what has been used here which then makes the MCMC analysis much slower. Hence computing the kSZ signal using the present code would require some further work on \texttt{SCRIPT}, possibly in the direction of making the code more efficient. 

In another different work, \texttt{SCRIPT} has also been extended to make detailed comparisons with the Ly$\alpha$ opacity fluctuations at $z \sim 5.5 - 6$ \citep{2021MNRAS.501.5782C}. Interestingly, these fluctuations indicate that the completion of reionization can be delayed up to $z \sim 5.5$. These constraints can have interesting consequences for our analysis, hence it becomes important to combine the extensions needed to model the Ly$\alpha$ opacity fluctuations with those presented in this paper. Unfortunately, this too faces some technical challenges as the modelling of the Ly$\alpha$ opacity requires additional calculations and introduction of more free parameters, both leading to slowing down of the code. These challenges are currently being worked on.

The present code will be quite useful when more observations become available. An example would be better estimates of the UVLF at $z \sim 6-7$, particularly at the faint end. This would allow us to probe the effects of feedback and possibly put more stringent constraints on the thermal history. Some of the degeneracies between parameters can also be removed with the upcoming 21~cm data from the reionization epoch. These data sets would ensure that the relevance of the present analysis remains intact in the near future.

\section*{Acknowledgements}

The authors acknowledge support of the Department of Atomic Energy, Government of India, under project no. 12-R\&D-TFR-5.02-0700.

\section*{Data availability}

A basic version of the code, which does not include the effects of recombinations and feedback on ionization maps, used in the paper is publicly available at \url{https://bitbucket.org/rctirthankar/script}. The data obtained from the extensions of the code and presented in this article will be shared on reasonable request to the corresponding author (BM).

\bibliographystyle{mnras}
\bibliography{thermal_mc} 

\appendix

\section{Comparison of the default model with two more alternatives}
\label{app:alternates}

\begin{figure*}
    \includegraphics[width=0.99\textwidth]{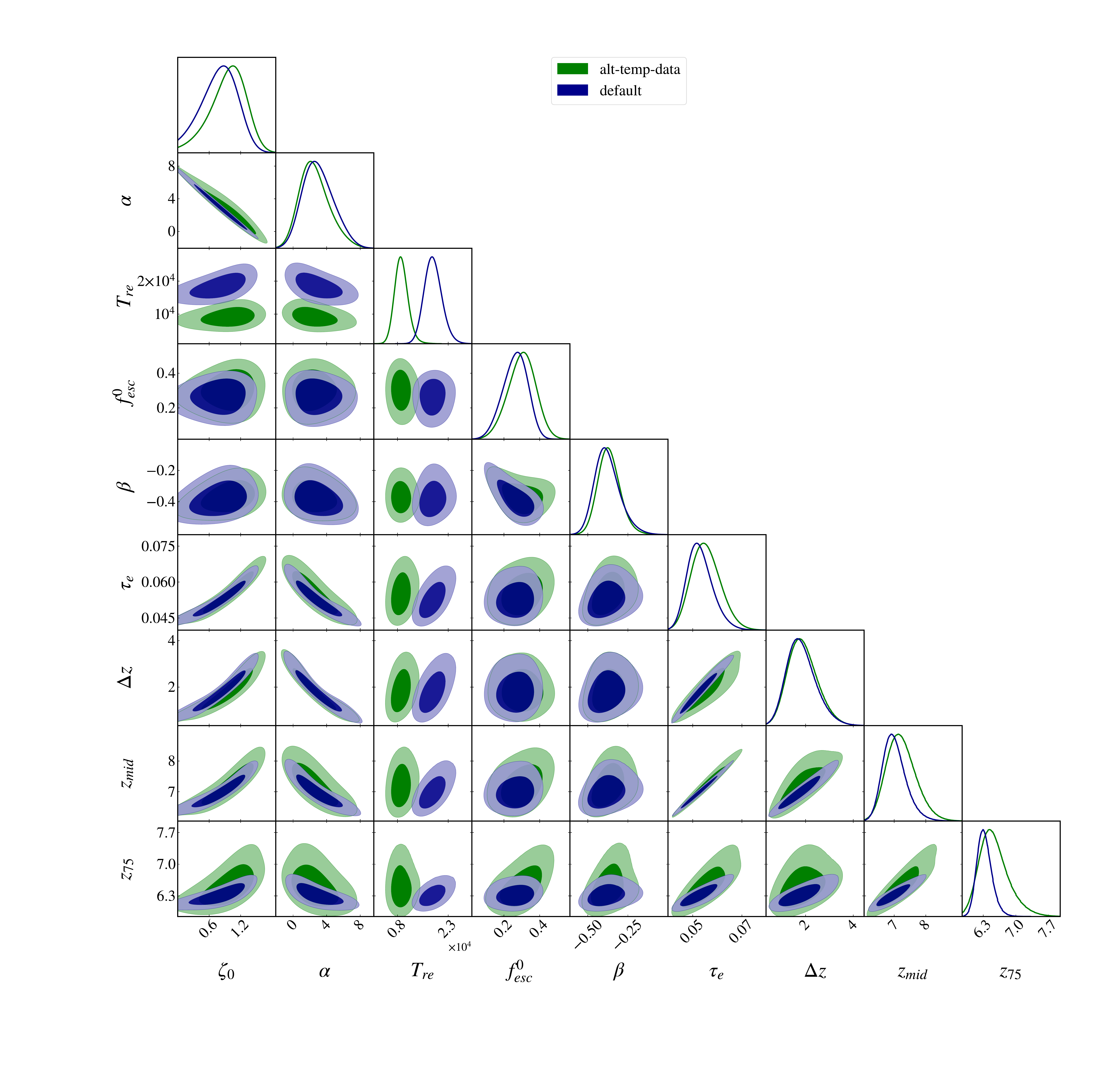}
    \caption{Posterior distributions of the five free and four derived parameters of our model for the default case and the alt-temp-data run where we use the \citet{2019ApJ...872...13W} temperature estimates instead of those from \citet{2020MNRAS.494.5091G}. In the diagonal panels, show the plots of marginalized one-dimensional posterior probabilities. The off-diagonal panels show the marginalized two-dimensional joint probability distributions. The contours represent the 68\% and 95\% confidence intervals.}
    \label{fig:corner_default_walther}
\end{figure*}

\begin{figure*}
    \includegraphics[width=0.99\textwidth]{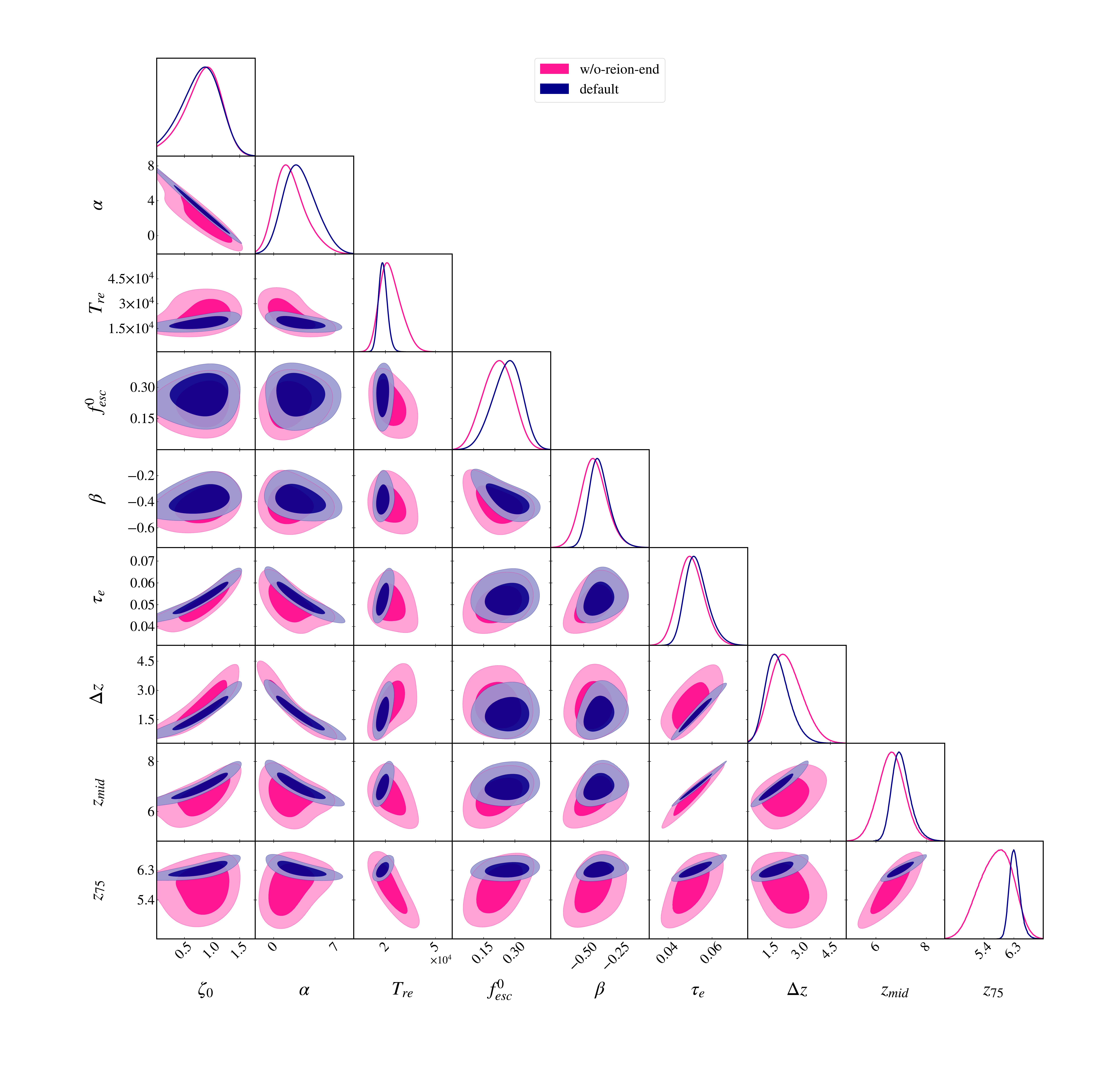}
    \caption{Posterior distributions of the five free and four derived parameters of our model for the default case and the w/o-reion-end run without the dark pixel constraints \citep{2011MNRAS.415.3237M} and the hard bound on reionization end. In the diagonal panels, we have shown the plots of marginalized one-dimensional posterior probabilities. The off-diagonal panels show the marginalized two-dimensional joint probability distributions. The contours represent the 68\% and 95\% confidence intervals.}
    \label{fig:corner_default_wo_darkpixel}
\end{figure*}

\begin{table*}
\caption{Parameter constraints obtained from the MCMC-based analysis for the default setup and two alternate cases. The first six rows correspond to the free parameters of the model while the others are the derived parameters. The free parameters are assumed to have uniform priors in the range mentioned in the second column. The other numbers are showing the mean value with 1$\sigma$ errors for different parameters using different combinations of MCMC runs. The bestfit values are quoted within the brackets for the different parameters. The value of $C_{\mathrm{HII}}$ is fixed to $3$ for all the cases.}
\begin{threeparttable}
\begin{tabular}{ccccc}
\hline
Parameters & Prior & default & alt-temp-data & w/o-reion-end \\
\hline
$\log(\zeta_0)$     & [$0,\infty$]     & $0.81^{+0.37}_{-0.32}~(0.92)$ & $0.97^{+0.38}_{-0.23}~(1.09)$ & $0.86^{+0.34}_{-0.27}~(0.90)$ \\ \\
$\alpha$ & [$-\infty,\infty$]  & $3.0^{+1.6}_{-2.1}~(2.3)$ & $2.5^{+1.2}_{-2.0}~(1.6)$ & $1.9^{+1.3}_{-1.9}~(2.3)$ \\ \\
$T_{\mathrm{re}}$ ($10^4$~K)    & [$0.01, 10$]      & $1.85^{+0.23}_{-0.27}~(1.83)$ & $ 0.91^{+0.16}_{-0.20}~(0.96)$ & $2.18^{+0.38}_{-0.61}~(1.79)$ \\ \\
$f_{\mathrm{esc}}^0$   & [$0,1$]       & $0.26^{+0.06}_{-0.06}~(0.36)$ & $0.30^{+0.08}_{-0.06}~(0.36)$    & $0.23^{+0.08}_{-0.07}~(0.34)$ \\ \\
$\beta$     &  [$-1,0$]         & $-0.38^{+0.07}_{-0.08}~(-0.44)$ & $-0.37^{+0.06}_{-0.07}~(-0.43)$  & $-0.41^{+0.09}_{-0.10}~(-0.46)$ \\ \\
\hline
Derived Parameters & & & &  \\ \\
$\tau_e$& &$0.053^{+0.004}_{-0.006}~(0.053)$ & $0.055^{+0.005}_{-0.006}~(0.056)$ & $0.051^{+0.005}_{-0.006}~(0.053)$  \\ \\
$\Delta z$& &$1.81^{+0.51}_{-0.67}~(1.87)$ & $1.87^{+0.46}_{-0.69}~(2.10)$ &  $2.20^{+0.65}_{-0.83}~(1.86)$\\ \\
$z_{\mathrm{mid}}$ &  &$7.00^{+0.30}_{-0.40}~(7.04)$ & $7.22^{+0.35}_{-0.51}~(7.20)$ & $6.69^{+0.51}_{-0.49}~(6.95)$\\ \\
$z_{\mathrm{75}}$ & &$6.32^{+0.14}_{-0.17}~(6.34)$ & $6.52^{+0.19}_{-0.35}~(6.45)$ &  $5.87^{+0.52}_{-0.40}~(6.26)$\\
\hline
\end{tabular}
\end{threeparttable}
\label{tab:param_cons_appendix}
\end{table*}

In this appendix, we show two more cases and compare the posteriors with our default setup. These are
\begin{itemize}
    
    \item \textbf{alt-temp-data:} There exists other measurements of the low density IGM temperature at $z=5.0~\&~5.4$, e.g., by \citet{2019ApJ...872...13W}. These estimates are very different than the estimates used for our default case, i.e., from \citet{2020MNRAS.494.5091G}. In particular, the value of $T_0$ at $z = 5.4$ in \citet{2019ApJ...872...13W} is almost half of that in \citet{2020MNRAS.494.5091G}. Hence using them together in the analysis will lead to poor fit between the model and data and may make the interpretation of the results more difficult. We rather carry out an analysis where we replace the default \citet{2020MNRAS.494.5091G} measurements with those of \citet{2019ApJ...872...13W}. The other observational constraints remain same.
    
    \item \textbf{w/o-reion-end:} As an interesting alternative, we explore a case where the constraints on the end stages of reionization, mostly derived from the quasar absorption spectra at $z \sim 5 - 6$, are relaxed. We thus drop the constraints on the global ionization fraction from dark pixel analysis \citep{2011MNRAS.415.3237M} and the hard prior on reionization end at $z=5.3$ in this case and run the MCMC analysis.
    
\end{itemize}

In Figure \ref{fig:corner_default_walther}, we show the posterior comparisons between our default model setup and the alt-temp-data case using \citet{2019ApJ...872...13W} temperature estimates. We find that these alternative data prefers a lower $T_{\mathrm{re}}$ value ($\approx 10^4~K$) compared to the default case. This is because the lower values of $T_0$ obtained by \citet{2019ApJ...872...13W}. Because of the lower $T_0$, relatively early reionization models are also acceptable by the data. The constraints on all the other parameters remain consistent with each other. The comparison statistics are shown in Table \ref{tab:param_cons_appendix}. Interestingly, one can find that the $T_{\mathrm{re}}$ estimates for our default setup show better agreement with the results from detailed simulation \citep{2019ApJ...874..154D}, while those for alt-temp-data are clearly lower.

In Figure \ref{fig:corner_default_wo_darkpixel}, we show the parameter space comparison between the default setup and the w/o-reion-end case which contains neither the constraints from dark pixel analysis nor the hard prior on reionization end at $z=5.3$. Not surprisingly, we find that the data allows very late reionization models (as evident from the distribution of $z_{\mathrm{mid}}$ and $z_{\mathrm{75}}$). However, the other constraints are consistent with the default setup.











\bsp	
\label{lastpage}
\end{document}